\documentclass{article}

\usepackage{PRIMEarxiv}

\usepackage[utf8]{inputenc} % allow utf-8 input
\usepackage[T1]{fontenc}    % use 8-bit T1 fonts
\usepackage{hyperref}       % hyperlinks
\usepackage{url}            % simple URL typesetting
\usepackage{booktabs}       % professional-quality tables
\usepackage{amsfonts}       % blackboard math symbols
\usepackage{nicefrac}       % compact symbols for 1/2, etc.
\usepackage{microtype}      % microtypography
\usepackage{lipsum}
\usepackage{fancyhdr}       % header
\usepackage{graphicx}       % graphics
\graphicspath{{media/}}     % organize your images and other figures under media/ folder

\usepackage{microtype}
\usepackage{graphicx}
\usepackage{subfigure}
\usepackage{subcaption}
\usepackage{multirow}
\usepackage{caption}
\usepackage{booktabs} % for professional tables

\usepackage{algorithm}
\usepackage{algorithmic}

\usepackage{amsmath}
\usepackage{amssymb}
\usepackage{mathtools}
\usepackage{amsthm}

\theoremstyle{plain}
\newtheorem{theorem}{Theorem}[section]

\newtheorem{lemma}[theorem]{Lemma}

\theoremstyle{definition}
\newtheorem{definition}[theorem]{Definition}

\theoremstyle{remark}

\title{Prediction Intervals for Individual Treatment Effects in a Multiple Decision Point Framework using Conformal Inference
}

\author{
  Swaraj Bose \\
  University of Michigan \\
   \And
  Walter Dempsey \\
  University of Michigan \\
}

\begin{document}
\maketitle

\begin{abstract}
Accurately quantifying uncertainty of individual treatment effects (ITEs) across multiple decision points is crucial for personalized decision-making in fields such as healthcare, finance, education, and online marketplaces. 
Previous work has focused on predicting non-causal longitudinal estimands or constructing prediction bands for ITEs using cross-sectional data based on exchangeability assumptions. We propose a novel method for constructing prediction intervals using conformal inference techniques for time-varying ITEs with weaker assumptions than prior literature. We guarantee a lower bound for coverage, which is dependent on the degree of non-exchangeability in the data. 
Although our method is broadly applicable across decision-making contexts, we support our theoretical claims with simulations emulating micro-randomized trials (MRTs) -- a sequential experimental design for mobile health (mHealth) studies. We demonstrate the practical utility of our method by applying it to a real-world MRT - the Intern Health Study (IHS).
\end{abstract}

% keywords can be removed

\section{Introduction}
\label{intro}

% The estimation of individual treatment effects (ITEs) in the potential outcomes framework \cite{rubin2005causal}, is a cornerstone of causal inference, aiming to uncover heterogeneity in treatment responses across individuals.
Estimating the causal effect of an intervention or treatment is a fundamental problem in many fields, including economics, medicine, and machine learning \cite{rubin1974estimating, imbens2015causal}. The Average Treatment Effect (ATE) serves as a cornerstone in causal inference, representing the average difference in outcomes between treated and untreated units across the entire population. While ATE provides a global measure of treatment impact, it often masks effect heterogeneity across different subgroups of the population \cite{varadhan2013estimation}. To address this limitation, researchers have turned to the Conditional Average Treatment Effect (CATE), which allows for a more nuanced understanding of how treatment effects vary across individuals depending on their characteristics. %is a deterministic function of the expected treatment effect, given an individual's covariates. CATE allows for a more nuanced understanding of how treatment effects vary across individuals with different characteristics and
Recent efforts have been made to estimate and perform inference on the CATE \cite{wager2018, kim2019estimating, jacob2021cate}. Previous research explored methodologies for predictive inference on individualized treatment effects using Bayesian methods \cite{hill2011bayesian} and Gaussian processes \cite{alaa2017bayesian}.

Both ATE and CATE operate at the group or subgroup level, aggregating effects over multiple individuals. In reality, the ultimate goal of causal inference is often to understand the Individual Treatment Effect (ITE), which represents the causal effect of the treatment for a single individual. 
Quantifying uncertainty associated with these predictions is crucial in areas like precision medicine \cite{blackstone2019precision}, education \cite{ferron2010estimating}, and targeted marketing \cite{guelman2014optimal, hitsch2024heterogeneous}.
% , where wrong actions may lead to serious consequences. 

% Precise knowledge of ITEs enables personalized decision-making and facilitates understanding of treatment effect variability, which is especially critical in domains such as healthcare, education, and social policy \cite{panofsky2009behavior,hekler2020precision}. 
% However, estimating ITEs reliably in complex, time-varying settings presents unique challenges, particularly when dealing with settings such as micro-randomized trials (MRTs) - a sequential experimental design for mobile health (mHealth) studies -, where participants are randomized to treatments at multiple decision points \cite{walton2018optimizing}. Previous research has extensively explored methodologies for predictive inference of ITEs using Bayesian methods \cite{hill2011bayesian} and Gaussian processes \cite{alaa2017bayesian}.

Conformal inference has emerged as a powerful framework for constructing distribution-free prediction intervals with finite-sample guarantees \cite{vovk2005algorithmic, shafer2008tutorial}. Conformal methods provide rigorous uncertainty quantification for predictive models, even in nonparametric settings, which allows us to use complicated ``black-box" machine learning models like random forests and neural networks. 
% For more on conformal methods, we refer the reader to a tutorial \cite{shafer2008tutorial} and the book \cite{vovk2005algorithmic}.
% Prediction intervals for ITE using conformal methods have been built in the past where separate intervals are built for the potential outcomes and combined to generate an interval for the ITE \cite{lei2021}.
In the potential outcomes framework \cite{rubin2005causal}, previous efforts to construct prediction intervals for ITEs using conformal methods have typically involved generating separate intervals for the potential outcomes, which are then combined to form an interval for the ITE \cite{ kivaranovic2020conformal, lei2021, jin2023sensitivity, schroder2024conformal}. Further efforts have been made to decouple the estimation of nuisance parameters from the conformal prediction procedure \cite{meta}. Prediction intervals for ITE have also been constructed in clustered data environments \cite{wang2024conformal}. However, these methods provide intervals in the cross-sectional setting and/or operate under the strong assumption of exchangeable data - exchangeability of calibration and test data, i.i.d. observations etc. %, which is a key limitation.

Conformal prediction methods have previously been applied to time series and longitudinal data, enabling uncertainty quantification for time-dependent outcomes 
% under weaker conditions of time-series exchangeability 
while accounting for temporal correlation \cite{lin2022conformal, lin2022conformal2, xu2023sequential, batra2023conformal}. It is unclear how these methods may be extended to account for ITEs or other causal estimands, requiring additional considerations not addressed in the cited works.

\textbf{Contributions.} Our goal is to construct prediction intervals for ITEs in longitudinal settings such as those encountered in mobile health (mHealth) studies. These settings typically involve repeated measurements on individuals, with temporal correlations arising due to the longitudinal nature of the data.
% These settings typically involve repeated measurements on individuals, exhibiting temporal dependence in the measurements. 
% This leads to challenges involving non-exchangeable data, making the naive assumption of exchangeability unrealistic. 
As a result, the assumption of exchangeable data is generally violated, rendering naive approaches based on exchangeability inappropriate.
In this paper, we address this issue and our contributions are three-fold - (1) we propose a novel theoretical framework to build prediction intervals for time-varying ITEs, (2) we derive a lower bound on the coverage of our prediction intervals that accounts for the lack of exchangeability in longitudinal data, (3) we empirically investigate the impact of different weighting schemes during the calibration step, demonstrating how robust weighting strategies can effectively reduce the coverage gap.

\subsection{Related work} 
\label{relatedwork}

% shorten significantly (not more than 2 lines per paper)
Causal inference has benefited from advancements in machine learning that transform causal effect estimation into a supervised learning problem. The concept of meta-learners for CATE estimation, introduced in \cite{kunzel2019metalearners}, was used to provide valid prediction intervals for ITEs \cite{meta} in the cross-sectional setting, under strong assumptions of exchangeable data. 
% The method decouples nuisance parameter estimation from the conformal procedure using a two-step outcome regression that enables %$the analyst to conduct 
% inference on the target estimand of interest (ITE) without directly using estimates of the potential outcomes. 
% Their method is not directly applicable to our current setting as it assumes cross-sectional data, exchangeability of calibration and test data and i.i.d. samples. Our proposal extends the procedure to the time-varying non-exchangeable setting under weaker assumptions.
% The ``beyond exchangeability'' framework \cite{BarberBE} extends non-causal conformal methods to accommodate non-exchangeable scenarios, broadening their applicability to real-world data contexts. They provide a general theory to quantify the coverage gap due to non-exchangeability. This framework does not outline how to incorporate causal estimands in the potential outcomes framework. Our proposal integrates these ideas into the time-varying ITE setting. 
Conformalized Quantile Regression (CQR) \cite{cqr} combines the flexibility of quantile regression with the robust uncertainty quantification of conformal inference and often produces tighter intervals than mean regression counterparts \cite{cqrcomp}. 
% CQR enables valid data-adaptive prediction intervals %while being data-adaptive to
% which account for the degree of observed heteroskedasticity. It has been shown empirically that quantile-regression based methods provide tighter intervals as compared to their mean regression counterparts \cite{cqr, cqrcomp}. 
% Our proposed method will leverage quantile regression.
% to generate tighter intervals for our time-varying ITEs.

% A related area of research is in clustered data environments. The estimation of ITEs often requires specialized approaches to account for within-cluster correlation. However, these approaches typically focus on static or cross-sectional data, where treatment effects are aggregated or stratified by clusters. For instance, \cite{wang2024conformal} estimate treatment effects in multi-level data assuming that cluster membership is fixed and that observations within clusters are exchangeable. In contrast, our work addresses time-varying data structures where exchangeability cannot be assumed even within individuals, as treatments and outcomes evolve dynamically over time.

The problem of non-exchangeable data has been tackled previously through several different approaches. 
% \cite{vovk2005algorithmic} propose using Mondrian methods, which divide the data into groups but still assume that data within a group are exchangeable. 
% The Weighted Conformal Prediction (WCP) framework \cite{tibshirani2019conformal} 
% addresses covariate shift between training and test distributions. The WCP framework has been extended to generate predictive distributions of ITEs \cite{jonkers2024conformal}, as well as 
% % to better inform dose-response relationships 
% for estimating optimal dosing strategies \cite{verhaeghe2024conformal}.
% Extensions to WCP to generate predictive distributions of ITE \cite{jonkers2024conformal} or to better inform dose-reponse relationships in an attempt to determine the optimal dose \cite{verhaeghe2024conformal} have been worked on recently.  
The ``beyond exchangeability'' framework \cite{BarberBE} focuses on working with non-exchangeable data and quantifies the loss in coverage when naively assuming exchangeability in the full conformal setup (transductive conformal inference \cite{vovk2005algorithmic}). It has been shown that the split conformal setup (inductive conformal inference \cite{vovk2005algorithmic}) is valid for a specific class of strictly stationary $\beta$-mixing processes, wherein the dependence in the data decreases with time \cite{oliveira2024split}. Despite addressing the exchangeability issue, the integration of existing work with causal estimands like ITE remains an open question. 

The Weighted Conformal Prediction (WCP) framework \cite{tibshirani2019conformal} 
addresses covariate shift between training and test distributions. The WCP framework has been extended to generate predictive distributions of ITEs \cite{jonkers2024conformal}, as well as 
% to better inform dose-response relationships 
for estimating optimal dosing strategies \cite{verhaeghe2024conformal}. Our work differs from the WCP literature \cite{tibshirani2019conformal, jonkers2024conformal, verhaeghe2024conformal} in several key aspects. Their approach relies on the covariate shift assumption, requiring an accurate estimation of the high-dimensional likelihood ratio to ensure valid coverage. Additionally, their method assigns weights to data points based on observed values to correct for the known distribution shift. In contrast, our approach uses fixed weights that are independent of specific data points, allowing it to account for unknown deviations from exchangeability, provided these deviations are small enough to maintain a low coverage gap. Notably, if no distribution shift is present and the data are truly exchangeable, their method offers no coverage guarantee, whereas ours continues to provide exact coverage.

\section{Problem Setup and Background}

\subsection{Motivating setting: Micro-randomized Trial (MRT)}
% All submissions must follow the specified format.
% Although this method can be used in any setting where we have multiple decision points, for the sake of illustration and simplicity, we will use terminology specific to MRTs from now onward. The major difference is using the word ``action'' instead of ``decision''. 

Mobile health (mHealth) \cite{rehg2017mobile} leverages the use of smartphones, wearables, and digital technologies to deliver healthcare interventions, monitor health outcomes, and promote wellness in real-time. Micro-randomized trials (MRTs) \cite{walton2018optimizing}, using sequential experimental designs, allow us to develop informed mobile health interventions. In an MRT, every participant is sequentially randomized to receive an intervention or treatment at numerous decision points (possibly even hundreds or thousands). Participants are closely monitored over time, with data on characteristics and treatment responses collected through self-reports or via sensors.
% Participants are monitored intensively over time, with measurements on participant characteristics and response to treatments collected through self-report or via sensors.

Previous research has focused on proximal effect estimation, which 
% refer to the immediate or short-term causal effects of a treatment administered at a specific decision point on subsequent outcomes measured shortly after. These effects 
%capture how a treatment influences near-term responses, typically within the same or next observation window, rather than long-term or cumulative outcomes. 
typically captures how a treatment influences responses in the same or next observation window rather than long-term or cumulative outcomes.
Proximal effects are particularly important in adaptive interventions, such as those used in mHealth studies, where the goal is to personalize treatment based on immediate feedback and evolving conditions. Building prediction intervals for the ITE for participants at the different decision points would enhance decision making. 

We subsequently introduce notation specific to MRT literature. 
% Let $(i,j)$ refer to the $i^{th}$ person at the $j^{th}$ decision point, $X_{ij}$ denote the observed characteristics and historical data until the $j^{th}$ decision point, $A_{ij}$, the binary action (decision) taken at the $j^{th}$ decision point, $\bar{A}_{ij}$, be the set of all actions until decision point $j$, $Y_{ij}$, the observed outcome at decision point j, $Y_{ij}(\bar{A}_{ij}, 0)$ and $Y_{ij}(\bar{A}_{ij}, 1)$, the potential outcomes if $A_{ij} = 0$ or $1$ respectively at decision point j, and $t_{ij}$, the tuple $(i,j)$. If we have $N$ people and $T$ decision points, we can consider the data to be of the format $(X_{ij}, A_{ij}, Y_{ij}, t_{ij}) \in \mathcal{X} \times \{0,1\} \times \mathbb{R} \times \mathcal{T}$, where $A_{ij} = \{0,1\}$, $\mathcal{T} \in \{ 1,2,3,...,N \} \times \{ 1,2,...,T \}$. $t_{ij} = (i, j)~~ \forall ~~ i \in \{ 1,2,3,..., N \}, j \in \{1,2, ..., T\}$. For a test unit $(X_{I,J}, Y_{I,J})$, where only $X_{I,J}$ is observed, our goal is to build a prediction interval/set for the ITE using the data so that for a pre-specified coverage level $1-\alpha$, we have 
% $$\mathbb{P}\left\{ \left(Y_{I,J}(\bar{A}_{I,J},1) - Y_{I,J}(\bar{A}_{I,J},0) \right) \in \hat{C}(X_{I,J})\right\} \geq 1 - \alpha$$
Let $A_{ij}$ denote the binary action (treatment), $Y_{ij}$ denote the observed outcome and $X_{ij}$ denote a subset of the observed characteristics
for the $i^{th}$ individual until and including the $j^{th}$ decision point. Let $\bar{A}_{ij}$ be the set of all actions \textbf{until and not including} decision point $j$, $Y_{ij}(\bar{A}_{ij}, 0)$ and $Y_{ij}(\bar{A}_{ij}, 1)$, the potential outcomes if $A_{ij} = 0$ or $1$ respectively. 

Let $[k] = \{  1,2,...,k \} $. If we have $N$ individuals and $T$ decision points, we can consider the data to be of the format $(X_{ij}, A_{ij}, Y_{ij}) \in \mathcal{X} \times \{0,1\} \times \mathbb{R}, ~ \forall ~ (i,j) \in [N] \times [T]$,  where $A_{ij} \in \{0,1\}$. For a test unit $(X_{I,J}, Y_{I,J})$, where only $X_{I,J}$ is observed, our goal is to build a prediction interval/set $(\hat{C}(X_{I,J}))$ for the ITE using the data so that for a pre-specified coverage level $1-\alpha$, we have 

\vspace{-5pt}

$$\mathbb{P}\left\{ \left(Y_{I,J}(\bar{A}_{I,J},1) - Y_{I,J}(\bar{A}_{I,J},0) \right) \in \hat{C}(X_{I,J})\right\} \geq 1 - \alpha$$

\subsection{Counterfactual inference}

A core difficulty in causal inference is the inability to observe counterfactual outcomes \cite{imbens2015causal}, which makes inference on ITEs very challenging. We state our proposal in the potential outcomes framework using the potential outcomes $Y_{ij}\left(\bar{A}_{ij}, 1\right) ~\text{and}~ Y_{ij}\left(\bar{A}_{ij}, 0\right)$ to build prediction intervals for ITE.  We require additional notation. Let $\pi(X_{ij}, \bar{A}_{ij})$ denote the propensity score $P(A_{ij} = 1 | X_{ij}, \bar{A}_{ij})$ and $\mu_a(X_{ij}, \bar{A}_{ij})$ denote the conditional average outcome for fixed treatment~$a \in \{0,1\}$.

To avoid building ITE intervals that depend on nuisance parameters ($\varphi = (\pi, \mu_0, \mu_1)$), conformal meta-learners have been used to construct prediction intervals for the target parameter \cite{meta}. Data are used to construct pseudo-outcomes ($\widetilde{Y}_\varphi$) and covariate pairs, with inference using the transformed dataset. Specifically, inverse-probability weighted (IPW) or the doubly-robust (DR) learners may be used to convert observed outcomes to pseudo-outcomes, as they satisfy a weaker set of conditions (as opposed to the X learner) and achieve required target coverage \cite{meta}. However, this was done in the cross-sectional setting and associated theory relied strongly on exchangeability assumptions. In multiple decision point frameworks, like an MRT, where there are repeated measurements for each individual, we cannot claim that we are in the i.i.d. or exchangeable domain. 

\subsection{Problem with exact coverage}

Previous work has shown that exact coverage is not theoretically guaranteed when the data are not exchangeable or i.i.d. \cite{BarberBE}. Thus, we propose a method that will have identical coverage guarantees to existing methods if the data are in fact exchangeable, and quantify the difference in coverage resulting from violation of the exchangeability assumption. %$which is dependent on the degree to which the exchangeability assumption is violated.
We define the coverage gap to be
\begin{align*}
\text{Coverage} \text{ gap} = 1 - \alpha - \mathbb{P}\left\{ \left(Y_{I,J}(\bar{A}_{I,J}, 1) - Y_{I,J}(\bar{A}_{I,J}, 0) \right) \in \hat{C}(X_{I,J})\right\}
\end{align*}

% In multiple decision point frameworks like an MRT, where there are repeated measurements for each individual, we cannot claim that we are in the i.i.d. or exchangeable domain. 

\section{Methodology}

% In this section, we will describe how we can extend the ideas presented by \cite{meta} to a longitudinal setup without the need for exchangeability in the data, a key limitation in the developed theory. We borrow ideas from \cite{BarberBE} where they provide near target coverage guarantees without the assumption of exchangeable data. Our method possesses the strengths of both worlds, conducting inference on the target estimand instead of the nuisance parameters and provides coverage guarantees disregarding notions of exchangeability.
In this section, we will describe how we can build prediction intervals for ITEs in a longitudinal setup without the need for data exchangeability. % a key limitation in the prior literature. 
% We borrow ideas from \cite{BarberBE} where they provide near target coverage guarantees without the assumption of exchangeable data. 
Our method possesses the strengths of conducting inference on the target estimand instead of the nuisance parameters and providing coverage guarantees without assuming exchangeability.

\subsection{Assumptions}
\label{sec:assumptions}
% We require very minimal assumptions and few regarding the data generating process.
% We assume the following which are very standard in MRT and ITE literature 
We make the following standard assumptions from causal inference \cite{lei2021, boruvka2018assessing}.

\textbf{1. Positivity}. $0<\pi(X_{ij}, \bar{A}_{ij}) = P(A_{ij}=1|X_{ij}, \bar{A}_{ij})<1, ~~ \forall ~ X_{ij} \in \mathcal{X}$

\textbf{2. Consistency}. For every decision point, $j \leq T, ~~Y_{ij} = Y(\bar{A}_{ij}, A_{ij})$, where $A_{ij} = 0$ or $1$, meaning that the observed value equals the corresponding potential outcome.

% \textbf{Unconfoundedness}.  $\left( Y_{ij}(0), Y_{ij}(1)  \right) \perp A_{ij} | X_{ij}$

\textbf{3. Sequential ignorability}. For every decision point $j \leq T$, $\{ Y_{i,j+t}(\bar{a}_{i,j+t}, 1), Y_{i,j+t}(\bar{a}_{i,j+t}, 0) \}_{t\geq 0}$ $\perp\!\!\!\perp A_{ij} ~|~X_{ij} , ~ \bar{A}_{ij}$, that is, all future potential outcomes are independent of $A_{ij}$ given the history, treatment assignment history and covariates observed up to decision point j.

We also assume that \textbf{the propensity score ($\pi(X_{ij}, \bar{A}_{ij})$) is known},
% In other words, we operate under the assumption of a known action assignment mechanism, which is a standard premise in most prior work \cite{tibshirani2019conformal,lei2021, meta} in this domain and should not be considered a limitation of our work.
which is a standard assumption in most of the causal conformal literature \cite{lei2021, meta, tibshirani2019conformal}.  Moreover, the MRT is an experimental design with known propensity score and therefore the assumption is satisfied in our motivating setting. In Section \ref{app:ass_caus_diag}, a causal diagram is provided to clarify the underlying assumptions and a notation guide is provided in Appendix \ref{app:notation} to help disentangle the notation.

\subsection{Proposed approach}
\label{sec:method}

We present the algorithm in a sequential manner, leveraging the split conformal method for constructing prediction intervals. For a broad overview about the split conformal method please refer to Appendix \ref{app:split_conformal}. A step-by-step summary of our proposed method can be found in Algorithm \ref{alg:ite_prediction}.

\subsubsection{Step 1: Data splitting}
\label{datasplit}

The training dataset ($D_{\text{tr}}$) is partitioned into 3 mutually exclusive sets: 

$D_\varphi$, used for estimation of nuisance parameters $\varphi = (\pi, \mu_0, \mu_1)$. These estimates will be used to create pseudo-outcomes ($\Tilde{Y}_{\varphi, ij}$) in the other two datasets.

$D_\text{model}$, employed to fit quantile regression models, $(\hat{q}_\text{lo}, \hat{q}_\text{hi})$, for predicting the lower and upper quantiles for $\Tilde{Y}_{\varphi, ij}$, respectively.

$D_\text{cal}$, to get prediction intervals using $(\hat{q}_\text{lo}, \hat{q}_\text{hi})$, and calculate the conformity scores ($V_\varphi$ ) 
% \begin{equation} \label{conf_score}   
% V_{\varphi}(X,\Tilde{Y}_{\varphi},\hat{q}_\text{lo}, \hat{q}_\text{hi}) = \text{max}\{ \hat{q}_\text{lo}(X) - \Tilde{Y}_{\varphi}, \Tilde{Y}_{\varphi} - \hat{q}_\text{hi}(X) \}
% \end{equation}

\subsubsection{Step 2: Estimation of nuisance parameters}
\label{nuisance}

% Here we get an estimate of the nuisance parameters $(\hat{\varphi})$. 
The dataset $D_\varphi$ is used to estimate the nuisance functions $\mu_a(X_{ij}, \bar{A}_{ij})$ for $a \in \{0,1\}$. Our algorithm is agnostic to model specification, affording greater analyst flexibility.
As in most related work in this domain, we assume a known propensity score ($\pi$).

% ; for instance, one could use a simple linear regression model or a complicated neural network to get estimates of the nuisance functions

% However, the theory holds even if one were to choose to estimate $\pi$. Again, any simple or complicated machine learning algorithm may be deployed to get an estimate, $\hat{\pi}$.

\subsubsection{Step 3: Conversion into pseudo-outcomes}
\label{ipwdr}

The pseudo-outcomes are created as a function of the estimates of the nuisance parameters ($\hat{\varphi}$) and the observed outcome $Y_{ij}$. We may use the following meta-learners to accomplish this step.
\begin{itemize}
    \item Inverse Probability Weighted (IPW) learner 
    \begin{equation*}
        \widetilde{Y}_{\varphi,ij} = \frac{A_{ij} - \hat{\pi}(X_{ij}, \bar{A}_{ij})}{\hat{\pi}(X_{ij}, \bar{A}_{ij})(1 - \hat{\pi}(X_{ij}, \bar{A}_{ij}))} Y_{ij}
    \end{equation*}
    \item Doubly-Robust (DR) learner
    \begin{align*}
        \widetilde{Y}_{\varphi,ij} =
        \frac{A_{ij} - \hat{\pi}(X_{ij}, \bar{A}_{ij})}{\hat{\pi}(X_{ij}, \bar{A}_{ij})(1 - \hat{\pi}(X_{ij}, \bar{A}_{ij}))} \left( Y_{ij} - \hat{\mu}_{A_{ij}}(X_{ij}, \bar{A}_{ij}) \right) 
         + \hat{\mu}_1(X_{ij}, \bar{A}_{ij}) - \hat{\mu}_0(X_{ij}, \bar{A}_{ij})
    \end{align*}
\end{itemize}
The data in $D_\text{model}$ and $D_\text{cal}$ are transformed into pseudo-outcome and covariate pairs , $(\widetilde{Y}_{\varphi,ij}, X_{ij})$.

\subsubsection{Step 4: Quantile regression}
\label{qr}

Quantile regression in conformal inference has been shown to create tighter intervals with similar coverage guarantees to their mean regression counterparts \cite{cqr, cqrcomp}. Our method is versatile in allowing the use of any quantile regression algorithm. We will show how a linear model,
% for predicting quantiles 
a random forest model, and a neural network for predicting quantiles perform empirically (in Section \ref{sec:sims}). We build quantile regression models - $\hat{q}_\text{lo}(X_{ij})$ and $\hat{q}_\text{hi}(X_{ij})$ to get conditional quantiles of the pseudo-outcomes ($\Tilde{Y}_{\varphi, ij}$) using the $D_\text{model}$ dataset which are used for predicting the lower and upper limits of the prediction interval, respectively. In our method, $\hat{q}_\text{lo}$ estimates the $\frac{\alpha}{2}$-th quantile, while $\hat{q}_\text{hi}$ estimates the $(1-\frac{\alpha}{2})$-th quantile.

Theoretical guarantees described in Section \ref{sec:theory} are also valid for intervals created using mean regression but may lead to wider intervals.

\subsubsection{Step 5: Calibrating Prediction Intervals}
\label{calibration}

We use $\hat{q}_\text{lo}$ and $\hat{q}_\text{hi}$ to predict the lower and upper limits of the prediction intervals for every data point ($X_{ij}$) in the dataset $D_\text{cal}$. 

\textbf{Choice of conformity score}. We define the conformity score as 
\begin{equation}
    % \label{conf.score}
    V_{\varphi,ij} = \text{max}\{ \hat{q}_\text{lo}(X_{ij}) - \Tilde{Y}_{\varphi,ij}, \Tilde{Y}_{\varphi,ij} - \hat{q}_\text{hi}(X_{ij}) \}
\end{equation}
This conformity score accounts for both undercoverage and overcoverage \cite{cqr}. For all data points in $D_\text{cal}$, we compute $V_{\varphi,ij}$'s. Again, we remind the reader that the theory is not dependent on the choice of conformity score and can be slightly modified to account for other conformity scores.

\subsubsection{Step 6: Prediction Interval for test data}
\label{predint}

Given a new test data point $X_{I,J}$, we create the following prediction for its ITE
% \begin{align}
%     \hat{C}(X_{I,J}) = \left[ \hat{q}_\text{lo}(X_{I,J}) 
%     - Q_{1-\alpha}\left( \sum_{(i,j) \in D_\text{cal}} \overset{\sim}w_{ij} \cdot\delta_{V_{\varphi,ij}} + \overset{\sim}{w}_\infty \cdot\delta_{+\infty} \right), \right. \nonumber \\
%     \left. \hat{q}_\text{hi}(X_{I,J}) 
%     + Q_{1-\alpha}\left( \sum_{(i,j) \in D_\text{cal}} \overset{\sim}w_{ij} \cdot\delta_{V_{\varphi,ij}} +  \overset{\sim}{w}_\infty \cdot\delta_{+\infty} \right) \right] \label{eq:prediction_interval}
% \end{align}
% \begin{align}
%     \hat{C}(X_{I,J}) = \Bigg[ \hat{q}_\text{lo}(X_{I,J}) 
%     - Q_{1-\alpha} \Bigg( \sum_{(i,j) \in D_\text{cal}} \overset{\sim}w_{ij} \cdot\delta_{V_{\varphi,ij}} \nonumber \\
%     + \overset{\sim}{w}_\infty \cdot\delta_{+\infty} \Bigg), \nonumber \\
%     \hat{q}_\text{hi}(X_{I,J}) 
%     + Q_{1-\alpha} \Bigg( \sum_{(i,j) \in D_\text{cal}} \overset{\sim}w_{ij} \cdot\delta_{V_{\varphi,ij}} \nonumber \\
%     + \overset{\sim}{w}_\infty \cdot\delta_{+\infty} \Bigg) \Bigg] \label{eq:prediction_interval}
% \end{align}
\begin{align}
    \hat{C}(X_{I,J}) = \Bigg[ \hat{q}_\text{lo}(X_{I,J}) 
    - \hat{Q}, ~~
    \hat{q}_\text{hi}(X_{I,J}) 
    + \hat{Q} \Bigg] 
    \label{eq:prediction_interval}
\end{align}
where, $w_{ij}$ is the weight assigned to 
% the $(i,j)$-th data point in
$V_{\varphi,ij}\text{'s}$, $\overset{\sim}{w}_{ij}$ is the normalized weight $\left(\overset{\sim}{w}_{ij} = \frac{w_{ij}}{\sum_{(i,j)\in D_\text{cal}} w_{ij} + w_{\infty}}\right)$. $\hat{Q}=Q_{1-\alpha}\left( \sum_{(i,j) \in D_\text{cal}} \overset{\sim}w_{ij} \cdot\delta_{V_{\varphi,ij}} +  \overset{\sim}{w}_\infty \cdot\delta_{\infty} \right)$ is the $(1-\alpha)$-th weighted quantile of the distribution of $(V_{\varphi,ij}\text{'s})$. $w_{\infty}$ is the weight assigned to a point mass at infinity $(\delta_{\infty})$. This weighting scheme allows flexibility and robustness in the face of model misspecification and changes in the data generating process - common in longitudinal studies.

A step-by-step broad overview of the method can be found in Algorithm \ref{alg:ite_prediction}.

\begin{algorithm}
\caption{Conformal Prediction Intervals for ITE}% \footnote{Refer to Section \ref{predint} for definitions of $w_{ij}$, $\overset{\sim}{w}_{ij}$ and $Q_{1-\alpha}(\cdot)$.}}
\label{alg:ite_prediction}
\begin{algorithmic}[1]
\REQUIRE Training dataset $\mathcal{D}_{\text{tr}}$, Test data point $X_{I,J}$, significance level $\alpha$, quantile regression algorithm $\mathcal{A}=(q_\text{lo}, q_\text{hi})$
\ENSURE Prediction interval for ITE: $\hat{C}_{I,J}(X_{I,J})$

\STATE Split $\mathcal{D}_{\text{tr}}$ into $\mathcal{D}_\varphi$, $\mathcal{D}_\text{model}$, and $\mathcal{D}_\text{cal}$
\STATE Use $\mathcal{D}_\varphi$ to estimate nuisance functions: $\hat{\varphi} = \{\hat{\pi}, \hat{\mu}_0, \hat{\mu}_1\}$
\STATE Transform $\mathcal{D}_\text{model}$ and $D_\text{cal}$ into pseudo-outcome/covariate pairs $\{(\widetilde{Y}_{\varphi,ij}, X_{ij})\}$ using $\hat{\varphi}$ and a meta-learner (DR/IPW)
\STATE Train quantile regression algorithm $\mathcal{A} = (q_\text{lo}, q_\text{hi})$ on $(\widetilde{Y}_{\varphi,ij}, X_{ij}) \in D_\text{model}$.
\STATE Get prediction intervals on $D_\text{cal}$ using $(\hat{q}_\text{lo}, \hat{q}_\text{hi})$
\STATE Calculate conformity scores ($V_{\varphi, ij}$) based on the created prediction intervals and the pseudo-outcomes $(\widetilde{Y}_{\varphi,ij}$) in $D_\text{cal}$.
\STATE Prediction interval for ITE for new test data point built using Equation \ref{eq:prediction_interval}.
% - \tiny
% \begin{align*}
%     \hat{C}(X_{I,J}) = \left[ \hat{q}_\text{lo}(X_{I,J}) 
%     - Q_{1-\alpha}\left( \sum_{(i,j) \in D_\text{cal}} \overset{\sim}w_{ij} \cdot\delta_{V_{\varphi,ij}} + \overset{\sim}{w}_\infty \cdot\delta_{+\infty} \right), \right. \\
%     \left. \hat{q}_\text{hi}(X_{I,J}) 
%     + Q_{1-\alpha}\left( \sum_{(i,j) \in D_\text{cal}} \overset{\sim}w_{ij} \cdot\delta_{V_{\varphi,ij}} +  \overset{\sim}{w}_\infty \cdot\delta_{+\infty} \right) \right]
% \end{align*}
% \normalsize

\end{algorithmic}
\end{algorithm}

%%%%%%%%%% END ALGORITHM %%%%%%%%%%%

\section{Theory}
\label{sec:theory}

% We are going to introduce some new notation and remind the reader of some notation used in previous Sections to enhance readability.
We define $[k] = \{ 1,2,..., k \}$.
Let the training data, $D_\text{tr}$, consist of N individuals and T decision points and the data be of the form $(X_{ij}, A_{ij}, Y_{ij}) \in \mathcal{X} \times \{0,1\} \times \mathbb{R}, ~~ \forall ~ (i,j) \in [N] \times [T]$.

% As we had defined previously, $X_{ij}$ are observed characteristics or covariates for individual $i$ until decision point $j$, $A_{ij}$, the binary action for individual $i$ at decision point $j$, $Y_{ij}$, the observed outcome for individual $i$ at decision point $j$, and $t_{ij}$ is the tuple $(i,j)$. Then we have $N\times T$ data points.
As we had defined previously, for individual $i$ at decision point $j$, let $X_{ij}$ be a subset of observed characteristics or covariates until and including decision point $j$, $A_{ij}$, the binary action, and $Y_{ij}$, the observed outcome. Then we have $N\times T$ data points. We define $\bar{A}_{ij}$ to be the sequence of all actions on individual $i$ \textbf{until but not including} decision point $j$.

Let the potential outcomes be $Y_{ij}( \bar{A}_{ij}, 1)$ and $Y_{ij}(\bar{A}_{ij}, 0)$ for person $i$ at decision point $j$. Let $(\mu_1,\mu_0)$ denote conditional average outcomes and propensity score be $\pi(X_{ij}, \bar{A}_{ij})=P(A_{ij}=1|X_{ij}, \bar{A}_{ij})$. The nuisance parameters are $\varphi = (\mu_0, \mu_1, \pi)$. We assume that the propensity score $\pi$ is known (that is, the action mechanism is known).
% Let $[k] = \{1,2...,k\}$.

The training data $D_\text{tr}$ is split into 3 mutually exclusive sets: $D_\varphi$, used for estimation of nuisance parameters $\varphi = (\pi, \mu_0, \mu_1)$, $D_\text{model}$, to fit quantile regression models $(\hat{q}_\text{lo}, \hat{q}_\text{hi})$ for the pseudo outcomes ($\Tilde{Y}_{\varphi, ij}$), $D_\text{cal}$, to get prediction intervals using $(\hat{q}_\text{lo}, \hat{q}_\text{hi})$, and get the conformity scores ($V_\varphi$ ) -
\begin{equation} \label{conf_score}   
V_{\varphi}(X,\Tilde{Y}_{\varphi},\hat{q}_\text{lo}, \hat{q}_\text{hi}) = \text{max}\{ \hat{q}_\text{lo}(X) - \Tilde{Y}_{\varphi}, \Tilde{Y}_{\varphi} - \hat{q}_\text{hi}(X) \}
\end{equation}

% Let the number of individuals in $D_\text{cal}$ be $n_c$ and decision points observed be T. \\
% Suppose we have a test data point $(X_{I,J}, Y_{I,J}), ~(I,J) \in \{1,2,...N,... \} \times \{ 
% 1,2,...,T,...\}$, $(I,J) \notin \{ 1,2,...,N  \} \times \{ 1,2,...,T \}$. $Z_{ij} = (X_{ij},\Tilde{Y}_{\varphi,ij})$, $Z = (Z_{11}, Z_{12},..., Z_{1T}, Z_{21}, Z_{22}, ..., Z_{2T}   ..., Z_{n_c,T-1}Z_{n_c,T}, Z_{I,J}), ~$
% $ Z^{(i,j)} = Z^{(i,j), (I,J)}  = (Z_{11}, Z_{12}, ..., Z_{i,j-1}, Z_{I,J}, Z_{i, j+1},$ $Z_{i+1,1},  ..., Z_{n_c,1}, Z_{n_c,2}, ..., Z_{n_c, T}, Z_{i,j})$,\\
% where $Z_{i, j-1} = Z_{i-1, T},$ if $j=1$ and $Z_{i, j+1} = Z_{i+1, 1},$ if $j=T$. In other words, $Z^{(i,j)}$ swaps indices $(i,j)$ with $(I,J)$.

Let $|D_\varphi| = n_\varphi$, $|D_\text{model}| = n_\text{m}$ and $|D_\text{cal}| = n_\text{c}$. Suppose we have a test data point $(X_{I,J}, Y_{I,J}), ~(I,J) \in \mathbb{N}\times\mathbb{N}$, but $(I,J) \notin [N] \times [T]$. 
% $Z_{ij} = (X_{ij},\Tilde{Y}_{\varphi,ij})$, $Z = (Z_{11}, Z_{12},..., Z_{1T}, Z_{21}, Z_{22}, ..., Z_{2T}   ..., Z_{N,T-1}Z_{N,T}, Z_{I,J}), ~$
% $ Z^{(i,j)} = Z^{(i,j), (I,J)}  = (Z_{11}, Z_{12}, ..., Z_{i,j-1}, Z_{I,J}, Z_{i, j+1},$ $Z_{i+1,1},  ..., Z_{N,1}, Z_{N,2}, ..., Z_{N, T}, Z_{i,j})$,\\
% where $Z_{i, j-1} = Z_{i-1, T},$ if $j=1$ and $Z_{i, j+1} = Z_{i+1, 1},$ if $j=T$. In other words, $Z^{(i,j)}$ swaps indices $(i,j)$ with $(I,J)$.
Let $Z_{ij} = (X_{ij},\Tilde{Y}_{\varphi,ij})$. $Z = (Z_{1:n_\text{c}, 1:T}, Z_{I,J}), ~$
% $ Z^{(i,j)} = Z^{(i,j), (I,J)}  = (Z_{11}, Z_{12}, ..., Z_{i,j-1}, Z_{I,J}, Z_{i, j+1},$ $Z_{i+1,1},  ..., Z_{N,1}, Z_{N,2}, ..., Z_{N, T}, Z_{i,j})$,\\
% where $Z_{i, j-1} = Z_{i-1, T},$ if $j=1$ and $Z_{i, j+1} = Z_{i+1, 1},$ if $j=T$. In other words, $Z^{(i,j)}$ swaps indices $(i,j)$ with $(I,J)$.
$ Z^{(i,j)} = Z^{(i,j), (I,J)}  = (Z_{11}, Z_{12}, ..., Z_{i,j-1}, Z_{I,J}, Z_{i, j+1}$ $Z_{i+1,1},  ..., Z_{n_\text{c},1}, Z_{n_\text{c},2}, ..., Z_{n_\text{c}, T}, Z_{i,j})$,
where $Z_{i, j-1} = Z_{i-1, T},$ if $j=1$ and $Z_{i, j+1} = Z_{i+1, 1},$ if $j=T$. In other words, $Z^{(i,j)}$ swaps indices $(i,j)$ with $(I,J)$.

The weights assigned to $V_{\varphi, ij} \in D_\text{cal}$ are $w_{ij}$ (fixed a priori) and normalized weights are defined by:
% \begin{align} \label{weights}
% \overset{\sim}{w}_{ij} & = \frac{ w_{ij}}{w_{11} + w_{12} + ... + w_{n_c,T-1} + w_{n_{c},T} + w_{\infty}}   , ~~~~~~~
% \overset{\sim}{w}
% _{+\infty} & = \frac{w_{\infty}}{w_{11} + w_{12} + ... + w_{n_{c}, T-1} + w_{n_{c}, T} + w_{\infty}}
% \end{align}

% The weights assigned to $V_{\varphi, ij} \in D_\text{cal}$ are $w_{ij}$ (fixed a priori) and normalized weights are defined by:
% \begin{align} \label{weights}
% \overset{\sim}{w}_{ij} & = \frac{ w_{ij}}{w_{11} + w_{12} + ... + w_{N,T-1} + w_{N,T} + w_{\infty}} \nonumber  \\
% \overset{\sim}{w}
% _{+\infty} & = \frac{w_{\infty}}{w_{11} + w_{12} + ... + w_{N, T-1} + w_{N, T} + w_{\infty}}
% \end{align}
\begin{align} \label{weights}
\overset{\sim}{w}_{ij}  = \frac{ w_{ij}}{\sum_{(i,j) \in [n_\text{c}] \times [T]}w_{ij} + w_{\infty}}  , ~~~~~~~
\overset{\sim}{w}
_{+\infty}  = \frac{w_{\infty}}{\sum_{(i,j) \in [n_\text{c}] \times [T]}w_{ij} + w_{\infty}}
\end{align}

Algorithm $\mathcal{A} = (\hat{q}_\text{lo}, \hat{q}_\text{hi})$ is defined as
$\mathcal{A} : \cup_{n \geq 0} (\mathcal{X} \times \mathbb{R} \times \mathcal{T})^n \rightarrow$ \{measurable functions $\hat{S}: \mathcal{X} \rightarrow \mathbb{R}$\}.

We first define the model $(\hat{q}_\text{lo}, \hat{q}_\text{hi}) = \mathcal{A}\left( \left( x_{ij}, y_{ij}, t_{ij} \right): (i,j) \in \mathcal{K} \right)$, where $\mathcal{K}= \{ (1,1), (1,2),...,(1,T), (2,1), (2,2), ..., (n_c,T-1), (n_c,T), (I,J)  \}$ 

We then define a conformity score vector $V_\varphi(z) \in \mathbb{R}^{(n_c\times T)+1}$, with entries $\left( V_\varphi \left( z \right) \right)_k = V_\varphi(z_{k})$, $k\in\mathcal{K}$. This is our mapping of a data sequence $z = ((z_{ij})) \in (\mathcal{X}\times \mathbb{R}^{(n_c\times T)+1}), (i,j) \in \mathcal{K})$ to a vector of conformity scores $V_\varphi(z)$.
To disentangle notation, we have provided a notation guide (Appendix \ref{app:notation}).

% We first define the model $(\hat{q}_\text{lo}, \hat{q}_\text{hi}) = \mathcal{A}\left( \left( x_{ij}, y_{ij} \right): (i,j) \in \mathcal{K} \right)$, where $\mathcal{K}= \left\{ [N] \times [T], (I,J)  \right\}$ .

% We then define a conformity score vector $V_\varphi(z) \in \mathbb{R}^{(N\times T)+1}$, with entries $\left( V_\varphi \left( z \right) \right)_k = V_\varphi(z_{k})$, $k\in\mathcal{K}$. This is our mapping of a data sequence $z = ((z_{ij})) \in (\mathcal{X}\times \mathbb{R}^{(N\times T)+1}), (i,j) \in \mathcal{K})$ to a vector of conformity scores $V_\varphi(z)$. 
% To disentangle notation, we have provided a notation guide (Appendix \ref{app:notation}).

% We use $\text{d}_\text{TV}(A,B)$ to denote the distributional distance between the distributions of A and B. This will help quantify the coverage gap due to dropping the assumption of exchangeability.

% Recall that the prediction interval is built as in Equation \ref{eq:prediction_interval}.

\begin{theorem}
\label{pseudo_coverage}
    \textbf{(Nonexchangeable split conformal prediction)}. Let $\mathcal{A} = (\hat{q}_\text{lo}, \hat{q}_\text{hi})$ be an algorithm that maps a sequence of pairs $(X_{ij}, \Tilde{Y}_{\varphi, ij})$
    % $((X_{i1}, Y_{i1}), (X_{i2}, Y_{i2}), t_i)$ 
    to a fitted function. Let $V_\varphi$ be the conformity score associated with the predictions, defined as $V_\varphi(\hat{q}_\text{lo}, \hat{q}_\text{hi}) = \text{max}\{ \hat{q}_\text{lo}(X) - \Tilde{Y}_{\varphi}, \Tilde{Y}_{\varphi} - \hat{q}_\text{hi}(X) \}$. Let $\text{d}_\text{TV}(.,.)$ denote the total variation distance (TVD; Definition \ref{app:dtv_def})
    % \cite{BarberBE}.
    % between the distributions of A and B.
    Then for a test data point $X_{I,J}$, and pre-specified significance level $\alpha$, the non-exchangeable split conformal method satisfies 
    \begin{align*}
        \mathbb{P}\left\{ \Tilde{Y}_{\varphi, I,J} \in \hat{C}(X_{I,J})\right\} \geq  1 - \alpha - \sum_{k \in \mathcal{K}} \overset{\sim}{w}_{k} \cdot \text{d}_{\text{TV}}(V_\varphi(Z),V_\varphi(Z^{k}))
    \end{align*}
\end{theorem}

% This is very similar to the theorem presented in \cite{BarberBE}. 
All proofs are provided in Appendix \ref{app:proofs}. Using Theorem \ref{pseudo_coverage}, we are able to establish a lower bound on our coverage guarantee for the transformed pseudo-outcomes. 
% It is important to note that the coverage gap, which is equal to $\sum_{k \in \mathcal{K}} \overset{\sim}{w}_{k} \cdot \text{d}_{\text{TV}}(V_\varphi(Z),V_\varphi(Z^{k}))$, can be reduced if the weights $\overset{\sim}{w}_{k}$ are monotonically decreasing with increasing TVDs. An ideal robust weighting scheme would rely on this idea to attain close to desired coverage levels, when exchangeability assumptions are violated in our longitudinal setting (Section \ref{sec:sim_chgpts}). If the data are indeed exchangeable, which would imply that $\text{d}_{\text{TV}}(V_\varphi(Z),V_\varphi(Z^{k})) = 0, ~\forall ~ k \in \mathcal{K}$, any choice of weights would give us exact coverage.
Next, we will show that the actual ITE lies within our prediction interval $\hat{C}(X_{I,J})$. 

% We start by introducing the concepts of stochastic dominance, as defined in \cite{meta}. 

% \begin{definition}
% \label{sd}
%     \textbf{(Stochastic dominance)}. Let $F$ and $G$ be two cumulative distribution functions. $F$ has first order stochastic dominance (FOSD) on $G$, $F \geq_{(1)} G$, iff $F(x) \leq G(x), ~ \forall x$, with strict inequality for some $x$. $F$ has second order stochastic dominance (SOSD) on $G$, $F \geq_{(2)} G$, iff $\int_{-\infty}^x [G(t) - F(t)]dt \geq 0, ~ \forall x$, with strict inequality for some $x$.
% \end{definition}

% \begin{definition}
% \label{cd}
%     \textbf{(Convex dominance)}. $F$ has monotone convex dominance (MCX) over $G$, $F \geq_{mcx} G$, iff $\mathbb{E}_{X\sim F}[\mu(X)] \geq \mathbb{E}_{X\sim G}[\mu(X)]$, for all non-decreasing convex functions $\mu: \mathbb{R} \rightarrow \mathbb{R}$
% \end{definition}

Let the true unobserved ITE be $Y_\text{ij,ITE} = Y_{ij}(\bar{A}_{ij},1) - Y_{ij}(\bar{A}_{ij},0)$. We define the oracle conformity scores ($V^{*}$) -
% \begin{align}
%     &V^{*}(X,Y(1),Y(0),\hat{q}_\text{lo}, \hat{q}_\text{hi}) = \\ &\text{max}\left\{ \hat{q}_\text{lo}(X) - \left(Y\left(1\right) - Y\left(0\right)\right), \nonumber \right. \left.\left(Y\left(1\right) - Y\left(0\right)\right) - \hat{q}_\text{hi}(X) \right\}
% \end{align}
\begin{align}
    V^{*}(X,Y_\text{ITE},\hat{q}_\text{lo}, \hat{q}_\text{hi}) = \text{max}\left\{ \hat{q}_\text{lo}(X) - Y_\text{ITE}, \nonumber \right. \left.Y_\text{ITE} - \hat{q}_\text{hi}(X) \right\}
\end{align}

Based on definitions of stochastic and convex dominance (Definition \ref{sd} and Definition \ref{cd}), we state Theorem \ref{realitecov}.

\begin{theorem}
\label{realitecov}
    Let $V_\varphi$ and $V^*$ be the conformity scores based on the observed data and the oracle, respectively. Let the underlying data be $(X_{ij}, A_{ij}, Y_{ij}(0), Y_{ij}(1))$ while the observed data be $(X_{ij}, A_{ij}, Y_{ij}),~~ \forall~(i,j) \in \mathcal{K}$. Let $\bar{A}_{I,J}$ be the sequence of all treatments received by participant $I$ until decision point $J$, and $k = (i,j) \in \mathcal{K}$. Then for a test data point $X_{I,J}$, $\exists~ \alpha \in (0,1)$, such that the pseudo-interval $\hat{C}(X_{I,J})$ constructed using the non-exchangeable split conformal method on the dataset $\mathcal{D} = \{ (X_{ij}, A_{ij}, Y_{ij}) \}_{i=1}^n$ satisfies
\begin{align*}
    \mathbb{P}&\left[ \left\{  Y_{I,J}\left( \bar{A}_{I,J},1 \right)  -  Y_{I,J}\left(\bar{A}_{I,J},0\right) \right\}  \in \hat{C}(X_{I,J})  \right]  \geq \mathbb{P}\left\{ \Tilde{Y}_{\varphi, I,J} \in \hat{C}(X_{I,J})\right\} \geq \\ & 1 - \alpha -\sum_{k\in\mathcal{K} \backslash (I,J)} \overset{\sim}{w}_k \cdot \text{d}_\text{TV}(V_\varphi(Z), V_\varphi(Z^k)), ~\forall ~ \alpha \in (0, \alpha^{*})  
\end{align*}
if at least one of the following stochastic ordering conditions hold: $(i)$  (First order stochastic dominance) $V_\varphi \geq_{(1)} V^*, (ii) ~\text{(Second order stochastic dominance)}~  V_\varphi \leq_{(2)} V^*$ and $(iii)~\text{(Monotone convex dominance)}~  V_\varphi \geq_{mcx} V^*$. Under condition $(i)$, we have $\alpha^* = 1$.
\end{theorem}

% It is important to note that although Theorem \ref{realitecov} is similar in flavor to the Theorem proposed by \cite{meta}, there is one critical distinction. We prove the theorem without the need for exchangeability, a key assumption in their proof. 

% % We would like to point out that
% Our coverage guarantee is marginal over the distribution of $\bar{A}_{ij}$ encompassing all possible treatment sequences leading up to the decision point.
% % or all possible sequences of treatments prior to the decision point.

% We also note that if at least one of the stochastic ordering conditions holds for our meta-learner, our coverage for the real ITE will at least match our pseudo-outcome coverage. 
% % we are guaranteed to have coverage, at least as much as our pseudo-outcome coverage, for the real ITE.

Theorem \ref{realitecov} states that if any of the stochastic ordering conditions hold, our created prediction intervals will be valid for the true ITE. Finally, we show that for the IPW and DR learners, convex dominance (Definition \ref{cd}) holds using Theorem \ref{mcx_theorem}.

\textbf{Remarks.} 
In exchangeable settings, conformal methods can guarantee exact coverage of \( (1 - \alpha) \). However, in our longitudinal, non-exchangeable setting, exact coverage cannot be guaranteed \cite{BarberBE}. Theorem~\ref{pseudo_coverage} establishes a lower bound on coverage. The discrepancy from exact coverage,
—referred to as the \textbf{\emph{coverage gap}}—
% or the \emph{coverage gap},
is given by 
$\sum_{k \in \mathcal{K}} \overset{\sim}{w}_{k} \cdot \text{d}_{\text{TV}}(V_\varphi(Z),V_\varphi(Z^{k}))$. Intuitively, assigning higher weights $(\overset{\sim}{w}_{k})$ to data points similar to the test point (lower TVD) and lower weights to dissimilar points (higher TVD) helps reduce this gap. In longitudinal settings such as MRTs, proximal outcomes are typically more strongly associated with recent observations, with this dependence decaying over time. This motivates a weighting scheme that emphasizes temporally closer observations. In Section~\ref{sec:sim_chgpts}, we empirically demonstrate that such temporally informed weighting can substantially reduce the coverage gap compared to naive equal weighting, in scenarios common in MRTs.

Theorem~\ref{pseudo_coverage} closely parallels Theorem 2 
% a result 
in Barber et al.\ (2023)~\cite{BarberBE}. Our result holds in the longitudinal setting for ITEs, giving a novel lower bound on the coverage of our pseudo-outcomes. 
% A similar to Theorem~\ref{realitecov} was given by Alaa et al.\ (2023)~\cite{meta}; however, they assume a cross-sectional, exchangeable framework. Our theory extends this result to the longitudinal, non-exchangeable setting for ITEs, and we provide a proof under these relaxed assumptions — constituting a novel theoretical contribution. Unlike Alaa et al. (2023) \cite{meta}, who rely on exchangeability and uniform rank arguments, our proof handles the longitudinal, non-exchangeable setting by leveraging a non-exchangeable coverage bound (Theorem 4.1) to ensure valid prediction intervals for ITEs.
A result similar to Theorem~\ref{realitecov} was given in Theorem 1 by Alaa et al.\ (2023)~\cite{meta}; however, they assume a cross-sectional, exchangeable framework, and their proof relied on exchangeability and uniform rank arguments. Our theory extends this result to the longitudinal, non-exchangeable setting for ITEs, and we provide a proof under these relaxed assumptions by leveraging a non-exchangeable coverage bound obtained from Theorem \ref{pseudo_coverage} — constituting a novel theoretical contribution. 

We develop theory for quantile regression methods with a specific conformity score. However, we want to reiterate that the theory is not dependent on specific models or scores. Rather, it can be very easily modified for mean regression methods as well as any conformity score.
% with slight modifications to the proofs and theorems.

% The goal is to build a prediction interval for the ITE of this new data point. Details of how the method works is provided in Section \ref{method}. 

\section{Simulations}
\label{sec:sims}

In this section, we demonstrate the performance of our method using simulation experiments. We consider $\alpha = 0.05$ in all cases.

% \subsection{Data generation}
% \label{datagen}
\textbf{Data generation.}
To generate the data, we consider a longitudinal variant of the example proposed in \cite{wager2018}, \cite{lei2021} and also used in \cite{meta}. We simulate data sequentially, generating information for individuals one at a time, which aligns naturally with the concept of MRTs. Unless specified otherwise, we use number of individuals/people, $N_\text{total}$ = 2000 for all our simulations, with a 75\%-25\% training and test data split. The training data ($D_\text{tr}$; N = 1500) is further broken down into thirds to get $D_\varphi, D_\text{model}, D_\text{cal}$. Covariates at the initial decision point were generated from an equicorrelated multivariate Gaussian distribution, while covariates at subsequent decision points were generated as functions of the previously observed covariates plus an independent error term. The propensity score was simulated from a logistic model. We considered both a linear outcome and a non-linear outcome (with periodic behavior), as a function of covariates, current and prior actions and autoregressive (AR) errors. In certain simulation settings, where we consider a changepoint, we modified treatment to have a negative effect post changepoint, as opposed to a positive effect pre changepoint. More rigorous details about data generation can be found in Appendix \ref{app:datagen}. Unless otherwise specified, simulations use a linear outcome, without the presence of changepoints.

\textbf{Nuisance parameter estimation. }
All covariates, the current action $A_t$ and the previous action $A_{t-1}$ are used to estimate the nuisance functions $\hat{\varphi} = (\hat{\mu}_1, \hat{\mu}_0)$; propensity score $\pi$ is assumed to be known. Unless specified otherwise, we use a linear model with an $L1$ penalty (LASSO) to estimate the nuisance parameters. In all cases where linear models are used we add interaction terms between all covariates and $A_t$ for estimation.

% \subsection{Quantile regression}
% \label{sim_qr}
\textbf{Quantile regression. }
We have used a linear model with an $L1$ penalty (LASSO) for prediction of quantiles. Detailed descriptions have been provided in Appendix \ref{app:qr}

% \subsection{Differentiating between two types of predictions}
% \label{sec:sim_preds}
% Predictions can be of two different types
\textbf{Two types of predictions. }
Since we are in a longitudinal setting, there are two different types of predictions that may be of interest - predictions for a new set of individuals (we refer to this as \textbf{predicting downward}) or predictions for existing individuals but at future decision points (we refer to this as \textbf{predicting outward}). A diagrammatic representation about our assumed data structure as well as distinctions between the two types of predictions have been included in Appendix \ref{app:sec_data_structure}.
% To learn more about our assumed data structure and a diagrammatic representation to distinguish between the two types of predictions, please see Appendix \ref{app:sec_data_structure}.

\subsection{Importance of weighting scheme}
\label{sec:sim_chgpts}

In theory, if the data are exchangeable, any choice of weights would satisfy our desired coverage (discussion of Theorem \ref{pseudo_coverage}). In practice, data rarely satisfy exchangeability due to temporal correlations and time-varying treatment effects. Our objective is to select a weighting scheme that remains robust to changes in the data-generating process over time and is able to maintain a low coverage gap.
% In reality, data are hardly exchangeable due to the presence of temporal correlations and variation in treatment effects over time. 
% This motivates the need to have a weighting scheme that is robust to unknown changes in the data generation process and maintains closer to target coverage levels.
% The goal is to choose a weighting scheme that is robust to changes in the data generation process over the course of time. %, which is common in our longitudinal setting.
% To this end, we conducted a synthetic data experiment to elucidate how the presence of a changepoint (common in longitudinal studies like MRTs) impacts our procedure and how choosing weights more intelligently could help reduce the coverage gap. 
To assess this, we use a synthetic data experiment with a changepoint—common in longitudinal settings such as MRTs—to evaluate how our method performs and whether more informed weighting strategies can mitigate the resulting coverage gap.
% To investigate this, we conducted a synthetic data experiment to examine how the presence of a changepoint—common in longitudinal settings such as micro-randomized trials (MRTs)—affects our procedure, and how more informed weighting strategies can help mitigate the resulting coverage gap.
We evaluate two instances in our linear outcome setup -

\textbf{Predicting outward}. We generate data on $T=90$ decision points but only use data until $T=30$ to train the models. We define the changepoint at $t_c=45$. In this case, we do not observe a changepoint in our training data, and this can be considered to be a case of model misspecification. As our test dataset, we use $D_\text{cal}$ with all decision points post 30. 

\textbf{Predicting downward}. We generate data on $T=520$ decision points and the algorithm is trained on all decision points in $D_\text{tr}$. We define the changepoint at $t_c=500$. In this case, we do observe a changepoint in our training data, but the data prior to changepoint constitutes more than 95\% of the training data. We use $D_\text{test}$ as our test dataset. 

\textbf{Choice of weights}. We show differences between our method's performance with respect to two choices of weights - (1) equal weights (E) assigned to all calibration scores $w_{ij} = 1,~ \forall i,j$, (2) decaying weights (D) assigned to calibration scores $w_{ij} = \psi^{|t-j|},~ \forall i,j$, where $t$ is the decision point for prediction. We set $\psi = 0.7$. Other weighting approaches as well as sensitivity analysis for the choice of $\psi$ have been provided in more detail in Appendix \ref{app:weig_schemes}.

\begin{figure}[ht]
\vskip 0.2in
\begin{center}
\centerline{\includegraphics[width=0.5\columnwidth]{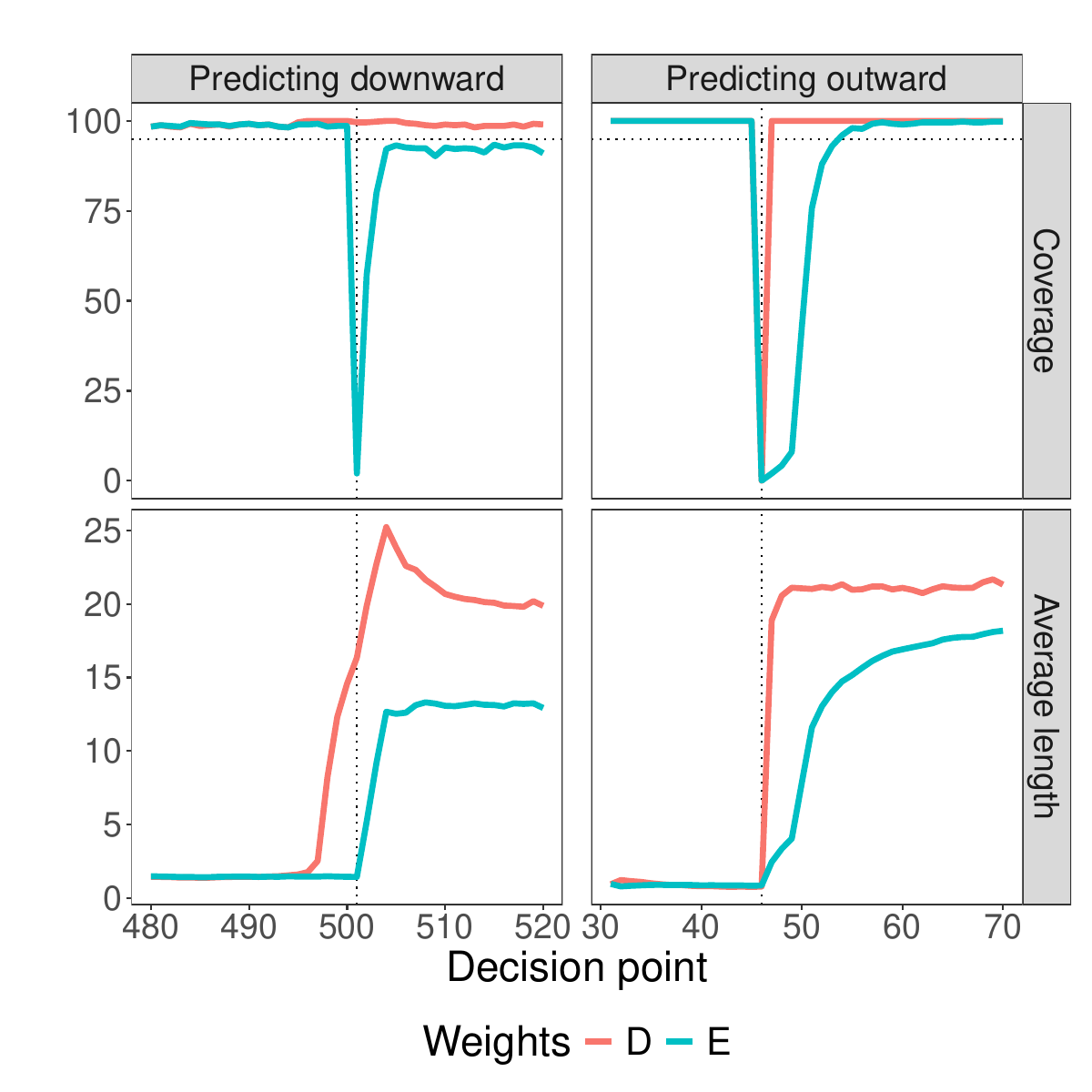}}
\caption{Comparison of algorithm performance when using two different weighting schemes - (i) Equal weights (E) and (ii) Decaying weights (D) - in the calibration step in the presence of changepoints. Horizontal dotted lines represent target coverage and vertical dotted lines represent the changepoints. 
% Plots have been cropped near their respective changepoints. 
More details about the two types of predictions can be found in Appendix \ref{app:sec_data_structure}. }
\label{fig:weights}
\end{center}
\vskip -0.2in
\end{figure}

Coverage (\%) and interval length are averaged over all 500 individuals at every decision point in the test dataset. From Figure \ref{fig:weights}, 
% we see that in both scenarios, the decaying weights method seems to be more robust to changepoints as compared to the naive equal weighting method. 
% while predicting downward, the decaying weights method is able to use information prior and post changepoint to ensure our target coverage is maintained by increasing interval length to account for the abrupt shift in action (treatment) effect. In contrast, the equal weights method is slower to recover and does not reach the desired target coverage level post changepoint. While predicting outward, the decaying weights method is faster to incorporate the newly observed changepoint information as it assigns higher weights to points that are closer temporally. Conversely, the equal weights method is slower to recover and has a converging trend towards the decaying weights method. 
when predicting downward, the decaying weights method utilizes information from both before and after the changepoint, adjusting interval lengths to preserve target coverage in response to abrupt shifts in treatment effect. In contrast, the equal weights method adapts more slowly and fails to reach the desired coverage post-changepoint. When predicting outward, the decaying weights method more rapidly incorporates changepoint information by prioritizing temporally proximate observations, whereas the equal weights method exhibits a slower recovery and gradually converges toward the performance of the decaying weights approach.

% \vspace{-25pt}

Selecting appropriate weights is a domain-specific challenge that often benefits from expert input and remains a promising direction for future research. 
% Choosing the weights properly is an interesting problem that is dependent on the domain of interest and experts' recommendations, and is an interesting area for future work. 
However, in settings like an MRT, where decision points are very frequent and the most recent outcome for an individual is highly correlated with their near future outcome(s), we suggest using decaying weights method (D), i.e., weights that decay temporally, as it is more robust to model misspecification and changes in the data generating process than the equal weights method. We will use the decaying weights method in all subsequent simulation setups and our real data example.

\subsection{Machine Learning (ML) models}
\label{sim_ml}

To showcase the flexibility of our algorithm to handle any machine learning model, we use a neural network with a single hidden layer and a random forest, and compare their performance with the linear model (LASSO). We note that the same machine learning model is used for both nuisance parameter estimation and quantile regression. We generate data for $T=50$ decision points and display results for predicting downward.
\\
From Table \ref{tab:ml_preddown}, we can see that irrespective of the setting (outcome type), all models are able to maintain target coverage levels. When we have a linear outcome, the linear model achieves performance comparable to the more complex machine learning models in terms of interval lengths. However, when we move to the setting with a non-linear outcome, the more complex machine learning models outperform the linear model and provide us with tighter prediction bands.

% \begin{tabular}{lrr}
% \toprule
% Model & Average length & Coverage\\
% \midrule
% Linear & 0.79 & 99.91\\
% Neural network & 0.81 & 95.77\\
% Random forest & \textbf{0.75} & 99.98\\
% \bottomrule
% \end{tabular}

% \begin{tabular}{lrr}
% \toprule
% Model & Average length & Coverage\\
% \midrule
% Linear & 3.42 & 100.00\\
% Neural network & 2.77 & 99.58\\
% Random forest & \textbf{2.30} & 100.00\\
% \bottomrule
% \end{tabular}

% \begin{table}[]
% \begin{tabular}{@{}lllll@{}}
% \toprule
% \multicolumn{1}{c}{\multirow{3}{*}{Model}} & \multicolumn{4}{c}{Outcome}                                 \\ \cmidrule(l){2-5} 
% \multicolumn{1}{c}{}                       & \multicolumn{2}{l}{Linear} & \multicolumn{2}{l}{Non-linear} \\
% \multicolumn{1}{c}{}                       & Coverage   & Avg. length   & Coverage     & Avg. length     \\ \cmidrule(r){1-1}
% Linear                                     & 99.91      & 0.79          & 100.00       & 3.42            \\
% Neural network                             & 95.77      & 0.81          & 99.58        & 2.77            \\
% Random Forest                              & 99.98      & 0.75          & 100.00       & 2.30            \\ \bottomrule
% \end{tabular}
% \end{table}

\begin{table}[ht]
\caption{Performance of ML models. Results are for predicting downward and are averaged over $N=500$ individuals in simulated test dataset. Best performing model's average length bolded. \emph{Abbreviations}: Cov, Coverage; PCov, Pseudo-outcome coverage; AL, Average length; LIN, Linear model with $L1$ penalty (LASSO); NN, Neural network; RF, Random forest.\\}
\centering
% \begin{tabular}{@{}lllll@{}}
% \toprule
% \multicolumn{1}{c}{\multirow{3}{*}{Model}} & \multicolumn{4}{c}{Outcome}                                 \\ \cmidrule(l){2-5} 
% \multicolumn{1}{c}{}                       & \multicolumn{2}{l}{Linear} & \multicolumn{2}{l}{Non-linear} \\
% \cmidrule{1-5}
% \multicolumn{1}{c}{}                       & Cov       & AL    & Cov         & AL      \\
% \cmidrule(r){1-5}
% LIN                                        & 99.91     & 0.79           & 100.00      & 3.42             \\
% NN                                         & 95.77     & 0.81           & 99.58       & 2.77             \\
% RF                                         & 99.98     & \textbf{0.75}           & 100.00      & \textbf{2.30}             \\ \bottomrule
% \end{tabular}
\small{
\begin{tabular}{@{}lllllll@{}}
\toprule
\multirow{2}{*}{} & \multicolumn{6}{c}{Outcome}                                                      \\ \cmidrule(l){2-7} 
                    Model   & \multicolumn{3}{c}{Linear} & \multicolumn{3}{c}{Non-linear} \\ \cmidrule(r){2-7}
                       & Cov         & PCov          & AL         & Cov          & PCov         & AL         \\
                       \cmidrule(r){1-7}
LIN                    & 99.91       & 95.10       & 0.79       & 100.00       & 95.14       & 3.42       \\
NN                     & 95.77       & 95.04       & 0.81       & 99.58       & 95.28       & 2.77       \\
RF                     & 99.98       & 94.90       & \textbf{0.75}       & 100.00       & 95.63       & \textbf{2.30}       \\ \bottomrule
\end{tabular}
}
\label{tab:ml_preddown}
%\end{centering}
\end{table}

\subsection{Pseudo-outcome coverage as surrogate for real ITE coverage}
\label{sec:pseudo_worstcase}

In Theorem \ref{realitecov}, we have seen that if certain conditions are met, real ITE coverage is at least as much as our computed pseudo-outcome coverage. Although the pseudo-outcomes created will differ with respect to the machine learning model used to estimate nuisance parameters, the result should hold irrespective of model choice. We verify this empirically, based on our simulations. From Table \ref{tab:ml_preddown} and Table \ref{tab:weight_preds} (Appendix \ref{app:add_results}),
% and sensitivity analysis performed in Section \ref{app:sensitivity}
we see that irrespective of the outcome type and model used, our real ITE coverage is always greater than our pseudo-outcome coverage. Additionally, for all sensitivity analyses (Appendix \ref{app:sensitivity_analysis}), we find that our claim is true. We are going to leverage this idea for our real data example, where coverage can never be computed in its truest sense.

\section{Real data example: Intern Health Study (IHS), 2018}

% \subsection{Intern Health Study (IHS), 2018}
% \subsection{Intern Health Study (IHS)}
\label{sec:ihs}

The Intern Health Study (IHS), 2018, is an MRT, where sleep and activity data were collected for 1563 medical interns over a period of 6 months between June and December 2018 \cite{necamp2020assessing}. The medical interns who were enrolled in the study were randomized to receive 4 kinds of notifications - activity, sleep, mood, or no notifications for a particular week (all with equal probability 1/4). Data were collected on a daily basis and aggregated at the weekly level. The three outcomes of interest were step count, sleep (minutes) and mood (measured on a scale of 1-10). A more detailed description can be found in Appendix \ref{app:ihs_desc}. We build prediction intervals for the ITE of each outcome separately. We consider the effect of assigning any treatment (action) versus none in a given week on each outcome. Prediction intervals were built for transformed outcomes - Step count $= \sqrt[3]{\text{Step count}}$ and Sleep $= \sqrt{\text{Sleep (minutes)}}$, while mood was kept on the original scale.

\begin{table}[ht]
\caption{Performance of ML models when creating prediction intervals for IHS data when predicting downward. Results averaged over all individuals and decision points in the test data. Best performing model's average length bolded. \emph{Abbreviations}: PCov, Pseudo-outcome coverage; AL, Average length; LIN, Linear model with $L1$ penalty (LASSO); NN, Neural network; RF, Random forest.\\}
\centering
{\small
\begin{tabular}{@{}lllllll@{}}
\toprule
\multirow{2}{*}{} & \multicolumn{6}{c}{Outcome}                                                      \\ \cmidrule(l){2-7} 
                    Model   & \multicolumn{2}{l}{Step count} & \multicolumn{2}{l}{Sleep} & \multicolumn{2}{l}{Mood} \\ \cmidrule(r){2-7}
                       & PCov         & AL          & PCov         & AL          & PCov         & AL         \\
                       \cmidrule(r){1-7}
LIN                    & 95.27       & 17.50       & 95.23       & \textbf{10.26}       & 94.64       & 5.73       \\
NN                     & 95.53       & \textbf{17.20}       & 95.28       & 12.01       & 94.43       & \textbf{5.60}       \\
RF                     & 95.25       & 17.61       & 95.39       & 10.41       & 95.04       & 5.87       \\ \bottomrule 
\end{tabular}
}

\label{tab:ihs_preddown}
% \end{centering}
\end{table}

Table \ref{tab:ihs_preddown} shows performance of different machine learning algorithms on a hold out test data subset (roughly 25\%) from IHS 2018, when predicting downward (averaged over all individuals and decision points/weeks). 
% Results for predicting outward can be found in Appendix \ref{app:ihs}. 
% It is important to note that 
Coverage is estimated based on created pseudo-outcomes, since we never observe true ITE in reality. All ML models achieve close to target levels of coverage.
% Appendix \ref{app:ihs} contains a detailed discussion of results.
A more detailed discussion of results has been included in Appendix \ref{app:ihs}.
\\Results of another real data experiment on IHS cohort from 2020 is included in Appendix \ref{app:ihscomp}).

\section{Discussion}
\label{sec:discussion}

We have developed a new method for building prediction intervals for time-varying ITEs with finite-sample guarantees. Theorems \ref{realitecov} 
% and \ref{mcx_theorem} 
provides a lower bound for coverage in our non-exchangeable data setting. Through simulations we have verified the empirical performance of the algorithm and the importance of choosing the weights of the calibration step intelligently. Our experiments highlight the flexibility of our method in incorporating any sophisticated machine learning technique, while maintaining target coverage levels. The real data example demonstrates the method's potential to build intervals for ITE with near-target coverage based on our surrogate pseudo-outcomes. 

% \section{Limitations}

We now address a few limitations of our work and point out potential directions to further advance ITE prediction intervals. 
% Although a common assumption in most of the ITE literature, assuming that the propensity score ($\pi$) is known 
% We are only able to provide coverage guarantees that are marginal across the distribution of all individuals and their covariates. It could be of interest to see whether conditional guarantees can be achieved. 
Although we are able to provide a lower bound on the coverage in the absence of the exchangeability assumption, this is dependent on the degree to which the assumption is violated. One could impose further restrictions on the class of data (stationary processes or $\beta$-mixing data) and reduce the coverage gap using developed theory 
% theory similar to 
\cite{oliveira2024split}. We note that the assumption of a known propensity score ($\pi$) is common throughout the literature and is not a limitation to our method, as it is satisfied in experimental designs like MRTs. However, future work could explore settings without a known propensity score. Finally, exact characterization of $\alpha^{*}$ in Theorem \ref{realitecov} is difficult \cite{meta} and beyond the scope of this paper, but could be an intriguing area for research.

\newpage
%Bibliography
\bibliography{main_arxiv}
\bibliographystyle{unsrt}

\newpage

\section{Appendix}

\subsection{Split conformal method}
\label{app:split_conformal}

Split conformal methods (also referred to as inductive conformal inference in \cite{vovk2005algorithmic}) have been widely utilized in prior research and applications \cite{oliveira2024split,vovk2020conformal,angelopoulos2022conformal, deutschmann2024adaptive}. 
% The idea is having a training dataset split into two parts - the model building dataset, used to train the model to provide predictions of the interval or associated measures, and the calibration dataset, used to calibrate these intervals bu adjusting for errors in prediction.
The approach broadly involves dividing the training dataset ($D_\text{tr}$) into two subsets: the model-building dataset ($D_\text{model}$), which is used to train the model and generate initial predictions or interval estimates, and the calibration dataset ($D_\text{cal}$), which is employed to refine these intervals by accounting for prediction errors using ``conformity scores''.

Let's consider a simple regression setup with a training dataset of the form $D_\text{tr} = (Y_i, X_i), ~ i=1,2,...,n$, which we split into two mutually exclusive subsets - $D_\text{model} = (Y_i, X_i), ~ i=1,2,...,n_1$, and $D_\text{cal} = (Y_i, X_i), ~ i=1,2,...,n_2$, $(n_1 +n_2 = n)$. 

\textbf{Mean regression}. We build a model $\hat{\mu}$ using $D_\text{model}$ and get estimates $\hat{\mu}(X_i)$ for all $(X_i)$ in $D_\text{cal}$. Next, we define a conformity score ($V(X,Y)$) associated with our predictions, for example $V_i = |Y_i - \hat{\mu}(X_i)|$. For a test point $(X_{n+1})$, a prediction interval is built as 
\begin{align*}
    \hat{C}(X_{n+1}) = \left[ \hat{\mu}(X_{n+1}) - Q_{(1-\alpha)}(V_i), \right. \\ 
    \left. \hat{\mu}(X_{n+1}) + Q_{(1-\alpha)}(V_i) \right]
\end{align*}
where $Q_{(1-\alpha)}(V_i)$ is the observed $(1-\alpha)$-th quantile of the distribution of $(V_1, V_2, ..., V_{n_2})$
\vspace{5pt}

\textbf{Quantile regression}. Here, we mainly focus on Conformalized Quantile Regression (CQR) \cite{cqr}. The basic workflow remains the same. We use $D_\text{model}$ to build quantile regression estimates $\hat{q}_\text{lo}$ and $\hat{q}_\text{hi}$ to estimate the lower and upper conditional quantiles (for instance the $\frac{\alpha}{2}$-th and $(1-\frac{\alpha}{2})$-th quantiles). Conformity scores are defined as $V_i = \text{max}\{ \hat{q}_\text{lo} - Y_i, Y_i - \hat{q}_\text{hi} \}$. Finally, for a test point $X_{n+1}$, a prediction interval is built as 
\begin{align*}
    \hat{C}(X_{n+1}) = \left[ \hat{q}_\text{lo}(X_{n+1}) - Q_{(1-\alpha)}(V_i),\right. \\
    \left. \hat{q}_\text{hi}(X_{n+1}) + Q_{(1-\alpha)}(V_i) \right]
\end{align*}
where $Q_{(1-\alpha)}(V_i)$ is the observed $(1-\alpha)$-th quantile of the distribution of $(V_1, V_2, ..., V_{n_2})$

\newpage

% \subsection{Detailed description of method}
% \label{app:method_details}

\newpage

\subsection{Theory}

\subsubsection{Assumptions and causal diagram}
\label{app:ass_caus_diag}

To disentangle some of the notation used in Section \ref{sec:theory} and to better elucidate the assumptions in Section \ref{sec:assumptions}, we present the following directed acyclyic graph (DAG).

\begin{figure}[ht]
    \centering
    \includegraphics[width=0.75\linewidth]{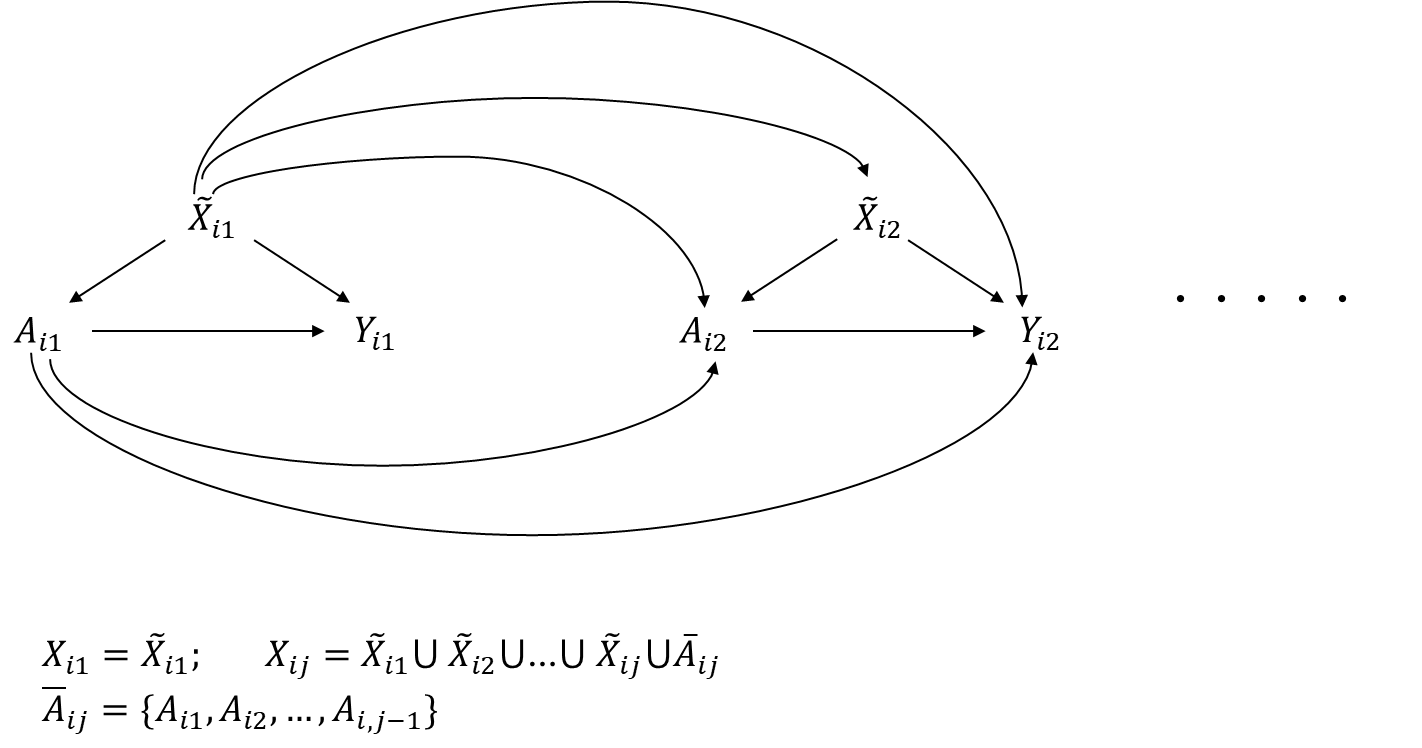}
    \caption{A two decision point illustration of the causal assumptions involved in the proposed method for a particular individual.}
    \label{fig:assumptions}
\end{figure}

We assume no interference and hence a particular individual can only affect their own characteristics and outcomes.

\subsubsection{Definitions}

\begin{definition}
\label{app:dtv_def}
    Consider a measurable space $(\Omega, \mathcal{F})$ and probability measures $P$ and $Q$ defined on $(\Omega, \mathcal{F})$. The total variation distance between P and Q is defined as 
    \begin{equation*}
        \text{d}_\text{TV}(P,Q) = \sup_{A \in \mathcal{F}} |P(A) - Q(A)|
    \end{equation*}
\end{definition}

\begin{definition}
\label{sd}
    \textbf{(Stochastic dominance)}. Let $F$ and $G$ be two cumulative distribution functions. $F$ has first order stochastic dominance (FOSD) on $G$, $F \geq_{(1)} G$, iff $F(x) \leq G(x), ~ \forall x$, with strict inequality for some $x$. $F$ has second order stochastic dominance (SOSD) on $G$, $F \geq_{(2)} G$, iff $\int_{-\infty}^x [G(t) - F(t)]dt \geq 0, ~ \forall x$, with strict inequality for some $x$.
\end{definition}

\begin{definition}
\label{cd}
    \textbf{(Convex dominance)}. $F$ has monotone convex dominance (MCX) over $G$, $F \geq_{mcx} G$, iff $\mathbb{E}_{X\sim F}[\mu(X)] \geq \mathbb{E}_{X\sim G}[\mu(X)]$, for all non-decreasing convex functions $\mu: \mathbb{R} \rightarrow \mathbb{R}$
\end{definition}

For all proofs, we use the full conformal framework on $D_\text{model}, D_\text{cal}$
% $D_\text{tr}$ 
and the test point $(X_{IJ}, Y_{I,J})$. Since split conformal is a special case of full conformal, all associated theory holds for our split conformal method as well. We proposed the split conformal variant in our main text as it is computationally more efficient than full conformal. We restate the theorems in the main text but in the full conformal setup.

\subsubsection{Full conformal framework}

\textbf{Notation.} While most of the notation remains the same we introduce the concept of data swapping, which is integral to the proof for full conformal inference. We mimic the setup described in Barber et al. (2023) \cite{BarberBE}. 

For our conformity scores using our quantile regression algorithm $\mathcal{A} = \{ \hat{q}_\text{lo}, \hat{q}_\text{hi} \}$, we define 

% \begin{equation}
%     \hat{S}_{y,k}(X,Y) = \mathcal{A} \left ( (X_{\pi_k (i,j)}, \Tilde{Y}^{y}_{\varphi, \pi_k (i,j)}, t_i): (i,j) \in \{ 1,2,...,n_c \} \times \{1,2,...,T\}  \right )
% \end{equation}

% where the permutation $\pi_k$ is the permutation on $\left( (1,1), (1,2), ..., (n_c,T-1), (n_c,T), (I,J)  \right)$ that swaps the indices $k$ and $(I,J)$, and where
% \begin{equation}
%     \Tilde{Y}_{\varphi,ij}^y = \Tilde{Y}_{\varphi,ij}, i=1,..,n_c,~~ j = 1,2,...,T  ~~ ; ~~ \Tilde{Y}^y_{\varphi,I,J} = y
% \end{equation}

\begin{equation}
    \hat{S}_{y,k}(X,\Tilde{Y}_\varphi) = \mathcal{A} \left ( (X_{\pi_k (i,j)}, \Tilde{Y}^{y}_{\varphi, \pi_k (i,j)}): (i,j) \in [n]\times [T]  \right )
\end{equation}

where $n = n_\text{c}+ n_\text{m}, |D_\text{cal}| = n_\text{c}, |D_\text{model}| = n_\text{m}$, $\pi_k$ is the permutation on $\left( (1,1), (1,2), ..., (n,T-1), (n,T), (I,J)  \right)$ that swaps the indices $k$ and $(I,J)$, and where
% \begin{equation}
%     \Tilde{Y}_{\varphi,ij}^y = \Tilde{Y}_{\varphi,ij},~~ i=1,..,n_c,~~ j = 1,2,...,T  ~~ ; ~~ \Tilde{Y}^y_{\varphi,I,J} = y
% \end{equation}
\begin{equation}
    \Tilde{Y}_{\varphi,ij}^y = \Tilde{Y}_{\varphi,ij},~~ (i,j) = [n] \times [T]  ~~ ; ~~ \Tilde{Y}^y_{\varphi,I,J} = y
\end{equation}

Define the scores from this model,
\begin{equation}
    V_{\varphi,ij}^{y,k} = V_\varphi\left(X_{ij}, \Tilde{Y}_{ij},\hat{S}_{y,k}(X_{ij},\Tilde{Y}_{ij})\right), (i,j) \in \mathcal{K} \backslash (I,J), ~~~ V^{y, k}_{\varphi,I,J} = V_\varphi\left(X_{ij}, \Tilde{Y}_{ij}, \hat{S}_{y,k}(X_{I,J}, y)\right)
\end{equation}
Then after drawing a random $K$ from a multinomial distribution where $K$ takes the value $(i,j)$ with probability $\overset{\sim}{w}_{ij}$, \\
\begin{equation}
    K \sim \sum^{}_{k \in \mathcal{K}} \overset{\sim}{w}_{k} \cdot \delta_{k}
\end{equation}
The prediction set is given by:
\begin{equation} \label{defn}
    \hat{C}(X_{I,J}) = \left\{ y \in \mathbb{R}: V^{y, K}_{\varphi,I,J} \leq Q_{1-\alpha} \left( \sum^{}_{k\in\mathcal{K}} \overset{\sim}{w}_k \cdot \delta_{V_{\varphi,k}^{y,K}} \right)  
 \right\}
\end{equation}

\subsubsection{Lemmas}

% \noindent \textbf{Lemma 1.} Let $Z_{ij} = (X_{ij}, Y_{ij})$, and $Z = (Z_{11}, Z_{12},...,Z_{1T}, Z_{21}, Z_{22},  ..., Z_{2T}, ..., Z_{n1},Z_{n2},...,Z_{nT},
% Z_{I,J})$. For any $k \in \mathcal{K} \left(\mathcal{K}= \{ (1,1), (1,2),...,(1,T), (2,1), (2,2), ..., (n,T-1), (n,T), (I,J)  \} \right)$, let $\pi_k$ denote the permutation on $\left( (1,1), (1,2), ..., (n,T-1), (n,T), (I,J)  \right)$ that swaps the indices $k$ and $(I,J)$. Suppose, 

\begin{lemma}
\label{lemma}
    Let $Z_{ij} = (X_{ij}, Y_{ij})$, and $Z = (Z_{1:n, 1:T}, Z_{I,J})$ and $\mathcal{K} = \{ [n]\times[T], (I,J)  \}$. For any $k \in \mathcal{K}$, let $\pi_k$ denote the permutation on $\left( (1,1), (1,2), ..., (n,T-1), (n,T), (I,J)  \right)$ that swaps the indices $k$ and $(I,J)$. Suppose, 
$$(V(Z^k))_{ij} = \text{max}\{ \hat{q}^k_\text{lo}(X_{\pi_k(i,j)}) - Y_{\pi_k(i,j)}, Y_{\pi_k(i,j)} - \hat{q}^k_\text{hi}(X_{\pi_k(i,j)}) \},$$
For any random $K$, where $K \sim \sum^{}_{k \in \mathcal{K}} \overset{\sim}{w}_{k} \cdot \delta_{k}$, let us define\\
\begin{equation*}
    (V(Z^K))_{ij} = 
\begin{cases} 
    V_{ij}^{Y_{I,J}, K}, ~~~ \text{if}~~ (i,j) \neq K ~~\text{and} ~~ (i,j) \neq (I,J) \\
    V_{I,J}^{Y_{I,J}, K}, ~~~ \text{if}~~ (i,j) = K \\
    V_{K}^{Y_{I,J}, K}, ~~~ \text{if}~~ (i,j) = (I,J)
\end{cases}
\end{equation*}

\noindent Then, we can verify deterministically that the definition of $(V(Z^K))_{ij}$ implies 
$$Q_{1-\alpha}\left( \sum^{}_{k\in\mathcal{K}\backslash (I,J)} \overset{\sim}{w}_{k} \cdot \delta_{V_{k}^{Y_{I,J}, K}} + \overset{\sim}{w}_{I,J} \cdot \delta_{+\infty}   \right) \geq Q_{1-\alpha}\left( \sum_{k \in \mathcal{K}}^{} \overset{\sim}{w}_k \cdot \delta_{(V(Z^K))_k} \right)$$

\noindent If $K=(I,J)$, then $V(Z^K)=V^{Y_{I,J},K}$, and so the bound holds trivially. If instead $K \neq (I,J)$, then the distribution on the left hand side equals
\begin{align*}
\sum_{k \in \mathcal{K} \backslash (I,J)}^{} \overset{\sim}{w}_k \cdot &\delta_{V_k^{Y_{I,J}, K}}~~ +~~ \overset{\sim}{w}_{I,J} \cdot \delta_{+\infty} ~~= \\ &\sum_{k \in \mathcal{K} \backslash (I,J)} \overset{\sim}{w}_k \cdot \delta_{V_k^{Y_{I,J}, K}} + \overset{\sim}{w}_K(\delta_{V_{K}^{Y_{I,J}, K}} + \delta_{+\infty}) + (\overset{\sim}{w}_{I,J} - \overset{\sim}{w}_K) \delta_{+\infty}
\end{align*}
The right hand side equals
$$\sum_{k\in\mathcal{K}}^{} \overset{\sim}{w}_k \cdot \delta_{(V(Z^K))_k} = \sum_{k\in\mathcal{K} \backslash (I,J)} \overset{\sim}{w}_k \cdot \delta_{V_k^{Y_{I,J}, K}} + \overset{\sim}{w}_K \cdot \delta_{V_{I,J}^{Y_{I,J}, K}} + \overset{\sim}{w}_{I,J} \cdot \delta_{V_{K}^{Y_{I,J}, K}}$$
$$= \sum_{k\in\mathcal{K} \backslash (I,J)} \overset{\sim}{w}_k \cdot \delta_{V_k^{Y_{I,J}, K}} + \overset{\sim}{w}_K(\delta_{V_{K}^{Y_{I,J}, K}} + \delta_{V_{I,J}^{Y_{I,J}, K}}) + (\overset{\sim}{w}_{I,J} - \overset{\sim}{w}_K) \delta_{V_{K}^{Y_{I,J}, K}}$$
Since $\overset{\sim}{w}_K \in [0, 1]$ by assumption, we have $\overset{\sim}{w}_{I,J} \geq \overset{\sim}{w}_K$. This verifies that the inequality holds. \\
\end{lemma}

\subsubsection{Theorems}

\begin{theorem}
\label{mcx_theorem}
    % Let $V(\hat{q}_\text{lo}, \hat{q}_\text{hi}) = \text{max}\{ \hat{q}_\text{lo}(X) - Y, Y - \hat{q}_\text{hi}(X) \}$ 
    Let the observed conformity scores be $V_\varphi(\hat{q}_\text{lo}, \hat{q}_\text{hi}) = \text{max}\{ \hat{q}_\text{lo}(X) - \Tilde{Y}_{\varphi}, \Tilde{Y}_{\varphi} - \hat{q}_\text{hi}(X) \}$ and the oracle conformity scores be $V^{*}(\hat{q}_\text{lo}, \hat{q}_\text{hi}) = \text{max}\{ \hat{q}_\text{lo}(X) - Y_\text{ITE}, Y_\text{ITE} - \hat{q}_\text{hi}(X) \}$
    and assume that the propensity score function $\pi : \mathcal{X} \rightarrow [0,1]$ is known. Then for any distribution $P(X, A, Y(0), Y(1))$, quantile regression estimates $\hat{q}_{\text{lo}}$ and $\hat{q}_{\text{hi}}$, nuisance estimate $\hat{\varphi}$, and transformed pseudo-outcome estimates $\widetilde{Y}_\varphi$, the IPW and DR-learners satisfy $V_{\varphi} \geq_{mcx} V^{*}$.
\end{theorem}

\newpage

\subsubsection{Proofs}
\label{app:proofs}

The proof of Theorem \ref{pseudo_coverage} is very similar to what one might encounter in Barber et al. (2023) \cite{BarberBE}. For the sake of completeness, we present it here. We state the full conformal version, but we note that the same coverage guarantees hold for split conformal prediction as well since it is a special case of full conformal. Our result holds in the longitudinal setting for ITEs, giving a novel lower bound on the coverage of our pseudo-outcomes.

\noindent\textbf{Theorem \ref{pseudo_coverage}} (Nonexchangeable full conformal prediction). Let $\mathcal{A} = (\hat{q}_\text{lo}, \hat{q}_\text{hi})$ be an algorithm that maps a sequence of pairs $(X_{ij}, \Tilde{Y}_{\varphi,ij})$
% $((X_{i1}, Y_{i1}), (X_{i2}, Y_{i2}), t_i)$ 
to a fitted function. Let $V$ be the conformity score associated with the predictions, defined as $V(\hat{q}_\text{lo}, \hat{q}_\text{hi}) = \text{max}\{ \hat{q}_\text{lo}(X) - \Tilde{Y}_\varphi, \Tilde{Y}_\varphi - \hat{q}_\text{hi}(X) \}$ and $\mathcal{K} = \{  [n]\times[T], (I,J) \}$. Then for a new test data point $X_{I,J}$, the non-exchangeable full conformal method satisfies 
$$\mathbb{P}\left\{ \Tilde{Y}_{\varphi,I,J} \in \hat{C}(X_{I,J})\right\} \geq 1 - \alpha - \sum_{k \in \mathcal{K}} \overset{\sim}{w}_k \cdot \text{d}_{\text{TV}}(V(Z),V(Z^{(i,j)}))$$\\
\begin{proof}

% Let $Z_{ij} = (X_{ij}, \Tilde{Y}_{\varphi,ij})$, and $Z = (Z_{11}, Z_{12},...,Z_{1T}, Z_{21}, Z_{22},  ..., Z_{2T}, ..., Z_{n1},Z_{n2},...,Z_{nT},
% Z_{I,J})$. For any $k \in \mathcal{K} \left(\mathcal{K}= \{ (1,1), (1,2),...,(1,T), (2,1), (2,2), ..., (n,T-1), (n,T), (I,J)  \} \right)$, as before, $\pi_k$ denotes the permutation on $\left( (1,1), (1,2), ..., (n,T-1), (n,T), (I,J)  \right)$ that swaps the indices $k$ and $(I,J)$. Then we can calculate for any $k \in \mathcal{K}$,
Let $Z_{ij} = (X_{ij}, \Tilde{Y}_{\varphi,ij})$, and $Z = (Z_{1:n, 1:T},
Z_{I,J})$. For any $k \in \mathcal{K}$, as before, $\pi_k$ denotes the permutation on $\left( (1,1), (1,2), ..., (n,T-1), (n,T), (I,J)  \right)$ that swaps the indices $k$ and $(I,J)$. Then we can calculate for any $k \in \mathcal{K}$,
\begin{equation}
    (V(Z^k))_{ij} = \text{max}\{ \hat{q}^k_\text{lo}(X_{\pi_k(i,j)}) - Y_{\pi_k(i,j)}, Y_{\pi_k(i,j)} - \hat{q}^k_\text{hi}(X_{\pi_k(i,j)}) \}
\end{equation}
Therefore for any random $K$, where $K \sim \sum^{}_{k \in \mathcal{K}} \overset{\sim}{w}_{k} \cdot \delta_{k}$,\\
\begin{equation} \label{v_cases}
    (V(Z^K))_{ij} = 
\begin{cases} 
    V_{ij}^{\Tilde{Y}_{\varphi,I,J}, K}, ~~~ \text{if}~~ (i,j) \neq K ~~\text{and} ~~ (i,j) \neq (I,J) \\
    V_{I,J}^{\Tilde{Y}_{\varphi,I,J}, K}, ~~~ \text{if}~~ (i,j) = K \\
    V_{K}^{\Tilde{Y}_{\varphi,I,J}, K}, ~~~ \text{if}~~ (i,j) = (I,J)
\end{cases}
\end{equation}

\noindent By the definition of nonexchangeable conformal prediction set from (\ref{defn}), we have 
\begin{equation}
    \Tilde{Y}_{\varphi,I,J} \notin \hat{C}\left(X_{I,J}\right) \iff V_{I,J}^{\Tilde{Y}_{\varphi,I,J}, K}>Q_{1-\alpha}\left(\sum_{k\in\mathcal{K}}^{} \overset{\sim}{w}_k \cdot \delta_{V_{k}^{Y_
{I,J}, K}}\right),
\end{equation}
and we can equivalently write this as
\begin{equation}
    \Tilde{Y}_{\varphi,I,J} \notin \hat{C}\left(X_{I,J}\right) \iff V_{ I,J}^{\Tilde{Y}_{\varphi,I,J}, K}>Q_{1-\alpha}\left(\sum_{k\in\mathcal{K} \backslash (I,J)}^{} \overset{\sim}{w}_k \cdot \delta_{V_{ k}^{\Tilde{Y}_{\varphi,I,J}, K}}+\overset{\sim}{w}_{I,J} \cdot \delta_{+\infty}\right) \text {. }
\end{equation}

\noindent Next, we have using \textbf{Lemma \ref{lemma}}, the definition of $(V(Z^K))_{ij}$ implies 
\begin{equation}
    Q_{1-\alpha}\left( \sum^{}_{k\in\mathcal{K}\backslash (I,J)} \overset{\sim}{w}_{k} \cdot \delta_{V_{k}^{\Tilde{Y}_{\varphi,I,J}, K}} + \overset{\sim}{w}_{I,J} \cdot \delta_{+\infty}   \right) \geq Q_{1-\alpha}\left( \sum_{k \in \mathcal{K}}^{} \overset{\sim}{w}_k \cdot \delta_{(V(Z^K))_k} \right)
\end{equation}

% \noindent If $K=(I,J)$, then $V(Z^K)=V^{\Tilde{Y}_{\varphi,I,J},K}$, and so the bound holds trivially. If instead $K \neq (I,J)$, then the distribution on the left hand side equals
% \begin{align*}
% \sum_{k \in \mathcal{K} \backslash (I,J)}^{} \overset{\sim}{w}_k \cdot &\delta_{V_k^{\Tilde{Y}_{\varphi,I,J}, K}}~~ +~~ \overset{\sim}{w}_{I,J} \cdot \delta_{+\infty} ~~= \\ &\sum_{k \in \mathcal{K} \backslash (I,J)} \overset{\sim}{w}_k \cdot \delta_{V_k^{\Tilde{Y}_{\varphi,I,J}, K}} + \overset{\sim}{w}_K(\delta_{V_{K}^{\Tilde{Y}_{\varphi,I,J}, K}} + \delta_{+\infty}) + (\overset{\sim}{w}_{I,J} - \overset{\sim}{w}_K) \delta_{+\infty}
% \end{align*}
% The right hand side equals
% $$\sum_{k\in\mathcal{K}}^{} \overset{\sim}{w}_k \cdot \delta_{(V(Z^K))_k} = \sum_{k\in\mathcal{K} \backslash (I,J)} \overset{\sim}{w}_k \cdot \delta_{V_k^{\Tilde{Y}_{\varphi,I,J}, K}} + \overset{\sim}{w}_K \cdot \delta_{V_{I,J}^{\Tilde{Y}_{\varphi,I,J}, K}} + \overset{\sim}{w}_{I,J} \cdot \delta_{V_{K}^{\Tilde{Y}_{\varphi,I,J}, K}}$$
% $$= \sum_{k\in\mathcal{K} \backslash (I,J)} \overset{\sim}{w}_k \cdot \delta_{V_k^{\Tilde{Y}_{\varphi,I,J}, K}} + \overset{\sim}{w}_K(\delta_{V_{K}^{\Tilde{Y}_{\varphi,I,J}, K}} + \delta_{V_{I,J}^{\Tilde{Y}_{\varphi,I,J}, K}}) + (\overset{\sim}{w}_{I,J} - \overset{\sim}{w}_K) \delta_{V_{K}^{\Tilde{Y}_{\varphi,I,J}, K}}$$
% Since $\overset{\sim}{w}_K \in [0, 1]$ by assumption, we have $\overset{\sim}{w}_{I,J} \geq \overset{\sim}{w}_K$. This verifies that the inequality holds. \\
\noindent Thus we have,
\begin{equation}
    \Tilde{Y}_{\varphi,I,J} \notin \hat{C}(X_{I,J}) \implies V_{I,J}^{\Tilde{Y}_{\varphi,I,J}, K} > Q_{1-\alpha} \left( \sum_{k\in\mathcal{K}} \overset{\sim}{w}_k \cdot \delta_{(V(Z^K))_k}  \right)
\end{equation}
Equivalently,
\begin{equation}
    \Tilde{Y}_{\varphi,I,J} \notin \hat{C}(X_{I,J}) \implies (V(Z^K))_K > Q_{1-\alpha} \left( \sum_{k\in\mathcal{K}} \overset{\sim}{w}_k \cdot \delta_{(V(Z^K))_k}  \right)
\end{equation}
Next, we define a function from $\mathbb{R}^{nT+1}$ to subsets $\mathcal{K}$ as follows. For any $r \in \mathbb{R}^{nT+1}$,
\begin{equation}
    \mathcal{S}(r) = \left\{ k \in \mathcal{K} : r_k > Q_{1-\alpha} \left( \sum_{k\in\mathcal{K}} \overset{\sim}{w}_k \cdot \delta_{r_k} \right)  \right\}
\end{equation}
These are the ``strange'' points - indices $k=(i,j)$ for which $r_k$ is unusually large, relative to the weighted empirical distribution of $r_{11}, r_{12}, ..., r_{nT}, r_{I,J}$. A direct argument (like in \cite{harrison2012conservative}) shows that 
\begin{equation}
    \sum_{k\in \mathcal{S}(r)} \overset{\sim}{w}_k \leq \alpha, ~~ \text{for all} ~~ r \in \mathbb{R}^{nT+1}
\end{equation}
In other words, the above means that the weighted fraction of ``strange'' points cannot exceed $\alpha$. Now,
\begin{equation}
    \Tilde{Y}_{\varphi,I,J} \notin \hat{C}(X_{I,J}) \implies K \in \mathcal{S}(V(Z^K))
\end{equation}
Finally,
\begin{align}
    \mathbb{P} \{ K \in \mathcal{S}(V(Z^K)) \} &= \sum_{k\in\mathcal{K}} \mathbb{P} \{ K=k ~~ \text{and} ~~ k \in \mathcal{S}(V(Z^k)) \} \nonumber \\
    &= \sum_{k\in\mathcal{K}} \overset{\sim}{w}_k \cdot \mathbb{P} \{ k \in S(V(Z^k)) \} \nonumber \\
    &\leq \sum_{k\in\mathcal{K}} \overset{\sim}{w}_k \left[\mathbb{P}\left\{k \in \mathcal{S}(V(Z))\right\} + \text{d}_{\text{TV}}\left(V(Z), V(Z^k)\right) \right] \nonumber \\
    &= \mathbb{E}\left[ \sum_{k \in \mathcal{S}(V(Z))} \overset{\sim}{w}_k  \right] + \sum_{k\in\mathcal{K} \backslash (I,J)} \overset{\sim}{w}_k \cdot \text{d}_\text{TV}(V(Z), V(Z^k)) \nonumber \\
    &\leq \alpha + \sum_{k\in\mathcal{K} \backslash (I,J)} \overset{\sim}{w}_k \cdot \text{d}_\text{TV}(V(Z), V(Z^k))
\end{align}

\end{proof} 
% This proves our theorem.

\newpage

% \noindent \textbf{Definition 1.1} \textbf{Stochastic dominance}. Let $F$ and $G$ be two cumulative distribution functions. $F$ has first order stochastic dominance (FOSD) on $G$, $F \geq_{(1)} G$, iff $F(x) \leq G(x), ~ \forall x$, with strict inequality for some $x$. $F$ has second order stochastic dominance (SOSD) on $G$, $F \geq_{(2)} G$, iff $\int_{-\infty}^x [G(t) - F(t)]dt \geq 0, ~ \forall x$, with strict inequality for some $x$.

% \noindent \textbf{Definition 1.2} \textbf{Convex dominance}. $F$ has monotone convex dominance (MCX) over $G$, $F \geq_{mcx} G$, iff $\mathbb{E}_{X\sim F}[\mu(X)] \geq \mathbb{E}_{X\sim G}[\mu(X)]$, for all non-decreasing convex functions $\mu: \mathbb{R} \rightarrow \mathbb{R}$

Although, the following proof looks similar in flavor to the proof in Alaa et al. (2023) \cite{meta}, there is one very important distinction that we remind the reader about. We prove the theorem for our longitudinal non-exchangeable data setting, whereas they had proved it under the assumption of exchangeability, in a cross-sectional setting. Their proof relied on exchangeability and uniform rank arguments. In contrast, we relax the assumption of exchangeability and use the non-exchanegable result from Theorem \ref{pseudo_coverage} to prove this theorem - constituting a novel theoretical contribution.

% \noindent \textbf{Theorem \ref{realitecov}.} Let $V_\varphi$ and $V^*$ be the conformity scores based on the observed data and the oracle, respectively. Let the underlying data be $(X_{ij}, A_{ij}, Y_{ij}(0), Y_{ij}(1))$ while the observed data be $(X_{ij}, A_{ij}, Y_{ij}), i = 1,...,n+1, ~~ j = 1,2,...,T ~~ k = (i,j) \in \mathcal{K} = \{ (1,1), (1,2),...(1,T), (2,1), (2,2), ..., (N,T-1), (N,T), (I,J)  \}$. Let $\bar{A}_{IJ}$ be the sequence of all treatments received by participant $I$ until decision point $J$. Then $\exists \alpha \in (0,1)$, such that the pseudo-interval $\hat{C}(X_{I,J})$ constructed using the non-exchangeable full conformal method on the dataset $\mathcal{D} = \{ (X_{ij}, A_{ij}, Y_{ij}) \}_{i=1}^n$ satisfies
% \begin{equation*}
%     \mathbb{P}\left[ \left\{Y_{I,J}(\bar{A}_{IJ},1) - Y_{I,J}(\bar{A}_{IJ},0)\right\} \in \hat{C}(X_{I,J})\right] \geq 1 - \alpha - \sum_{k\in\mathcal{K} \backslash (I,J)} \overset{\sim}{w}_k \cdot \text{d}_\text{TV}(V_\varphi(Z), V_\varphi(Z^k)), ~ \forall ~ \alpha \in (0, \alpha^{*})  
% \end{equation*}
% if at least one of the following stochastic ordering conditions hold: $(i) V_\varphi \geq_{(1)} V^*, (ii) V_\varphi \leq_{(2)} V^*$ and $(iii) V_\varphi \geq_{mcx} V^*$. Under condition $(i)$, we have $\alpha^* = 1$.
% \\ 
% \begin{proof}

\noindent \textbf{Theorem \ref{realitecov}.} Let $V_\varphi$ and $V^*$ be the conformity scores based on the observed data and the oracle, respectively. Let the underlying data be $(X_{ij}, A_{ij}, Y_{ij}(0), Y_{ij}(1))$ while the observed data be $(X_{ij}, A_{ij}, Y_{ij}),~~ \forall~(i,j) \in \mathcal{K}$. Let $\bar{A}_{I,J}$ be the sequence of all treatments received by participant $I$ until decision point $J$, and $k = (i,j) \in \mathcal{K}$. Then for a test data point $X_{I,J}$, $\exists~ \alpha \in (0,1)$, such that the pseudo-interval $\hat{C}(X_{I,J})$ constructed using the non-exchangeable full conformal method on the dataset $\mathcal{D} = \{ (X_{ij}, A_{ij}, Y_{ij}) \}_{i=1}^n$ satisfies
\begin{align*}
    \mathbb{P}\left[ \left\{  Y_{I,J}\left( \bar{A}_{I,J},1 \right)  -  Y_{I,J}\left(\bar{A}_{I,J},0\right) \right\}  \in \hat{C}(X_{I,J})  \right]  &\geq \mathbb{P}\left\{ \Tilde{Y}_{\varphi, I,J} \in \hat{C}(X_{I,J})\right\} \geq 1 - \alpha - \\ 
    &\sum_{k\in\mathcal{K} \backslash (I,J)} \overset{\sim}{w}_k \cdot \text{d}_\text{TV}(V_\varphi(Z), V_\varphi(Z^k)), ~ \forall ~ \alpha \in (0, \alpha^{*})  
\end{align*}
if at least one of the following stochastic ordering conditions hold: $(i) V_\varphi \geq_{(1)} V^*, (ii) V_\varphi \leq_{(2)} V^*$ and $(iii) V_\varphi \geq_{mcx} V^*$. Under condition $(i)$, we have $\alpha^* = 1$.
\\ 
\begin{proof}

Recall that the pseudo-interval was constructed as all points belonging to the set (\ref{defn}): \\
% \begin{equation*}
%     \hat{C}(X_{n+1}) = [\hat{q}_\text{lo}(X_{n+1}) - Q_{1-\alpha}(V_\varphi, \mathcal{D}_\text{cal}), \hat{q}_\text{hi}(X_{n+1}) + Q_{1-\alpha}(V_\varphi, \mathcal{D}_\text{cal})]
% \end{equation*}
$$\hat{C}(X_{I,J}) = \left\{ y \in \mathbb{R}: V^{y, K}_{\varphi,I,J} \leq Q_{1-\alpha} \left( \sum_{k\in\mathcal{K}} \overset{\sim}{w}_k \cdot \delta_{V_{\varphi,i}^{y,K}} \right)  
 \right\}$$
% where, $Q_{1-\alpha}(V_{\varphi}, \mathcal{D}_\text{cal}) := \begin{cases}
%     V_{\varphi,\ceil*{(n_c + 1)(1-\alpha)(1+\frac{1}{n_c})}}, ~~~~ ~~~~~~ \alpha \geq \frac{1}{n_c + 1} \\
%     \infty, ~~~~~~~~~~~~~~~~~~~~~~~~~~~~~~~~~~~ o.w.
% \end{cases}
% $ \\ \\
Note that the two events are equivalent
\begin{equation}
    \left\{ \left(Y_{I,J}(\bar{A}_{IJ},1) - Y_{I,J}(\bar{A}_{IJ},0)\right) \in \hat{C}(X_{I,J}) \right\} \iff \left\{ V_{I,J}^* \leq Q_{1-\alpha} \left( \sum_{k\in\mathcal{K} } \overset{\sim}{w}_k \cdot \delta_{V_{\varphi,k}^{y,K}} \right)  
  \right\}
\end{equation}

\noindent Consider the data $(X_{ij}, \Tilde{Y}_{\varphi,ij}, t_{ij}), i = 1,...,n, ~~ j = 1,2,...T$. Using Theorem \ref{pseudo_coverage}, we have:
\begin{align}
    & \mathbb{P}\left\{V^{y, K}_{\varphi,I,J} > Q_{1-\alpha} \left( \sum_{k\in\mathcal{K}} \overset{\sim}{w}_k \cdot \delta_{V_{\varphi,k}^{Y_{I,J}, K}} \right) \right\}  \leq \alpha + \sum_{k\in\mathcal{K} \backslash (I,J)} \overset{\sim}{w}_k \cdot \text{d}_\text{TV}(V_\varphi(Z), V_\varphi(Z^k)) \nonumber \\
    \implies & \mathbb{P}\left\{V^{y, K}_{\varphi,I,J} \leq Q_{1-\alpha} \left( \sum_{k\in\mathcal{K}} \overset{\sim}{w}_k \cdot \delta_{V_{\varphi,k}^{Y_{I,J}, K}} \right) \right\}  \geq 1 - \alpha - \sum_{k\in\mathcal{K} \backslash (I,J)} \overset{\sim}{w}_k \cdot \text{d}_\text{TV}(V_\varphi(Z), V_\varphi(Z^k))
\end{align}

\noindent Subsequently, the three conditions are examined. \\

\textbf{(i) FOSD} $\mathbf{V_\varphi \geq_{(1)} V^*}$ \\
If $V_\varphi \geq_{(1)} V^*$, then from Definition 1.1, $F_{V_\varphi}(v) \leq F_{V^*}(v), \forall~v.$ Equivalently, FOSD can be written as 
\begin{equation}
    V_\varphi \geq_{(1)} V^* \iff \mathbb{P}(V_\varphi \leq v) \leq \mathbb{P}(V^* \leq v)
\end{equation}
Using the above, we get
\begin{align}
    V_\varphi \geq_{(1)} V^* \implies \mathbb{P}\left(V_{I,J}^* \leq Q_{1-\alpha} \left( \sum_{k\in\mathcal{K}} \overset{\sim}{w}_k \cdot \delta_{V_{\varphi,k}^{y,K}} \right)\right) &\geq \mathbb{P}\left(V_{\varphi, I,J} \leq Q_{1-\alpha} \left( \sum_{k\in\mathcal{K}} \overset{\sim}{w}_k \cdot \delta_{V_{\varphi,k}^{y,K}} \right)\right) \nonumber\\
    &\geq 1 - \alpha - \sum_{k\in\mathcal{K} \backslash (I,J)} \overset{\sim}{w}_k \cdot \text{d}_{\text{TV}}(V_\varphi(Z), V_\varphi(Z^k))
\end{align}
Note that the above holds for any value of $\alpha$, hence $\alpha^* = 1$. This concludes the statement. \\

% \vspace{0.2cm} 
\textbf{(ii) SOSD} $\mathbf{V_\varphi \leq_{(2)} V^*}$ \\
If $V_\varphi \leq_{(2)} V^*$, then we know that $\exists~\alpha^* \in (0,1)$, where $F_{V_\varphi}^{-1}(\alpha^*) = F_{V^*}^{-1}(\alpha^*) = v^*$ and $F_{V^*}(v) \geq F_{V_\varphi}(v), \forall ~ v \geq v^*$. Equivalently, SOSD can be written as 
\begin{equation}
    V_\varphi \leq_{(2)} V^* \implies \exists ~ v^*, s.t.~~ \mathbb{P}(V_\varphi \leq v) \leq \mathbb{P}(V^* \leq v), ~\forall~v \geq v^* 
\end{equation}
Hence the following holds for all $\alpha \leq \alpha^*$
\begin{align}
    V_\varphi \leq_{(2)} V^* \implies \mathbb{P}\left(V_{I,J}^* \leq Q_{1-\alpha} \left( \sum_{k\in\mathcal{K}} \overset{\sim}{w}_k \cdot \delta_{V_{\varphi,k}^{y,K}} \right)\right) &\geq \mathbb{P}\left(V_{\varphi, I,J} \leq Q_{1-\alpha} \left( \sum_{k\in\mathcal{K}} \overset{\sim}{w}_k \cdot \delta_{V_{\varphi,k}^{y,K}} \right)\right) \nonumber \\
    &\geq 1 - \alpha - \sum_{k\in\mathcal{K} \backslash (I,J)} \overset{\sim}{w}_k \cdot \text{d}_{\text{TV}}(V_\varphi(Z), V_\varphi(Z^k))
\end{align}
for all $0\leq\alpha\leq \alpha^*$. This concludes (ii). \\

\textbf{(iii) MCX} $\mathbf{V_\varphi \geq_{mcx} V^*}$ \\
The proof is identical to (ii).\\

\end{proof}

\newpage

The proof of Theorem \ref{mcx_theorem} is very similar to that in the Appendix of Alaa et al. (2023) \cite{meta}, but we include it here for completeness. We are going to abuse notation and drop subscripts $(i,j)$ for the following proof.

\noindent \textbf{Theorem \ref{mcx_theorem}.} Let $V(\hat{q}_\text{lo}, \hat{q}_\text{hi}) = \text{max}\{ \hat{q}_\text{lo}(X) - Y, Y - \hat{q}_\text{hi}(X) \}$ and assume that the propensity score function $\pi : \mathcal{X} \rightarrow [0,1]$ is known. Then for any distribution $P(X, A, Y(0), Y(1))$, quantile regression estimates $\hat{q}_{\text{lo}}$ and $\hat{q}_{\text{hi}}$, nuisance estimate $\hat{\varphi}$, and transformed pseudo-outcome estimates $\widetilde{Y}_\varphi$, the IPW and DR-learners satisfy $V_{\varphi} \geq_{mcx} V^{*}$.\\

\begin{proof}

Let us define $Y_\pi$ and $Y_{1-\pi}$ for the IPW- and DR-learners
\begin{align}
    Y_\pi(x) &= \begin{cases}
        \frac{1}{\pi(x)}Y(1), ~~~~~~~~~~~~~~~~~~~~~~~~~~~~~~~~~~~~~~~~~~~~~~~ \text{for IPW-learner}  \\
        \frac{1}{\pi(x)} \left\{ Y(1) - \hat{\mu}_1(x) \right\}  + \hat{\mu}_1(x) - \hat{\mu}_0(x), ~~~~~~~~~~ \text{for DR-learner}
    \end{cases} \\
    Y_{1-\pi}(x)&= \begin{cases}
        \frac{-1}{1-\pi(x)}Y(0), ~~~~~~~~~~~~~~~~~~~~~~~~~~~~~~~~~~~~~~~~~~~~ \text{for IPW-learner} \nonumber \\
        \frac{-1}{1-\pi(x)} \left\{ Y(0) - \hat{\mu}_0(x) \right\}  + \hat{\mu}_1(x) - \hat{\mu}_0(x), ~~~~~~~ \text{for DR-learner}
    \end{cases}
\end{align}
We note that for both the IPW- and DR-learners,
\begin{equation}
    \pi(x)Y_\pi(x) + (1-\pi(x))Y_{1-\pi}(x) = Y(1) - Y(0)
\end{equation}
Let $V_\varphi(x)$ denote the conditional random variable $V_\varphi|X=x$, the conformity score evaluated at a given feature point. Similarly, $V^*(x)$ denotes the conditional random variable $V^*|X=x$, the oracle score evaluated at a given feature point. $V_\varphi(x)$ can be written as
\begin{equation*}
    V_{\varphi}(\hat{q}_\text{lo}, \hat{q}_\text{hi}) = \text{max}\{ \hat{q}_\text{lo}(X) - Y, Y - \hat{q}_\text{hi}(X) \}
\end{equation*}
can also be written as
\begin{align}
    V_{\varphi}(\hat{q}_\text{lo}, \hat{q}_\text{hi}) &= \frac{\hat{q}_\text{lo}(X) - \hat{q}_\text{hi}(X)}{2} + \frac{|\hat{q}_\text{lo}(X) + \hat{q}_\text{hi}(X) - 2\widetilde{Y}_\varphi|}{2} \nonumber\\
    &= \frac{\hat{q}_\text{lo}(X) - \hat{q}_\text{hi}(X)}{2} + \left| \frac{\hat{q}_\text{lo}(X) + \hat{q}_\text{hi}(X) }{2} - \widetilde{Y}_\varphi \right|
\end{align}
Similarly, the oracle scores can be written as:
\begin{equation}
    V^*(\hat{q}_\text{lo}, \hat{q}_\text{hi}) = \frac{\hat{q}_\text{lo}(X) - \hat{q}_\text{hi}(X)}{2} + \left| \frac{\hat{q}_\text{lo}(X) + \hat{q}_\text{hi}(X) }{2} - (Y(1) - Y(0)) \right|
\end{equation}

\noindent We note that $\widetilde{Y}_\varphi = f(A,Y),$ where $Y = \left( Y(0), Y(1) \right)$ \\

\noindent Let $\delta_{q_-} = \frac{\hat{q}_\text{lo}(X) - \hat{q}_\text{hi}(X)}{2}$ and $\delta_{q_+} = \frac{\hat{q}_\text{lo}(X) + \hat{q}_\text{hi}(X)}{2}$, and let $u$ be a non-decreasing convex function. Then we have
\begin{align}
    \mathbb{E}[u(V_\varphi(\hat{q}_\text{lo}, \hat{q}_\text{hi}))] &= \mathbb{E}_{A, Y}\left[ u \left( \delta_{q_-} + \left|   \delta_{q_+} - \widetilde{Y}_\varphi \right| \right)  \right] \nonumber \\
    &= \pi(x) \mathbb{E}_{Y}[u(\delta_{q_-} + |\delta_{q_+} - \widetilde{Y}_\pi|)] + (1-\pi(x))\mathbb{E}_{Y}[u(\delta_{q_-} + |\delta_{q_+} - \widetilde{Y}_{1-\pi}|)] \nonumber \\
    &\geq \mathbb{E}_{Y}[u(\pi(x)\delta_{q_-} + \pi(x) |\delta_{q_+} - \widetilde{Y}_\pi| + (1-\pi(x)) \delta_{q_-} + (1-\pi(x)) |\delta_{q_+} - \widetilde{Y}_{1-\pi}| )] \nonumber \\
    & \text{(due to convexity of $u$)} \nonumber \\
    & \geq \mathbb{E}_{Y}[u( \delta_{q_-} + |\delta_{q_+} - \pi(x) \widetilde{Y}_\pi + (1-\pi(x))\widetilde{Y}_{1-\pi}| )] \nonumber \\
    & \text{(triangle inequality and non-decreasing)} \nonumber \\
    &= \mathbb{E}_{Y}\left[u\left( \delta_{q_-} + |\delta_{q_+} - \left(Y\left(1\right) - Y\left(0\right)\right)| \right)\right] \nonumber \\
    &= \mathbb{E}_{A,Y}[u(V^*(\hat{q}_\text{lo}, \hat{q}_\text{hi}))] = \mathbb{E}[u(V^*(\hat{q}_\text{lo}, \hat{q}_\text{hi}))]
\end{align}
This implies that $V_\varphi \geq_{mcx} V^*$.

\end{proof}
\newpage

\subsection{Simulations}
\label{app:sim_settings}

All simulations were performed in R (version 4.2.2).

\subsubsection{Data generation}
\label{app:datagen}

To generate the data, we consider a longitudinal variant of the examples proposed in the prior literature \cite{wager2018, lei2021,meta}.
% \cite{wager2018} \cite{lei2021} and also used in \cite{meta}.
We simulate data sequentially, generating information for individuals one at a time, which aligns naturally with the concept of MRTs. Unless specified otherwise, we use number of individuals/people, $N_\text{total}$ = 2000 for all our simulations, with a 75\%-25\% training and test data split. The training data ($D_\text{tr}$; N = 1500) is further broken down into thirds to get $D_\varphi, D_\text{model}, D_\text{cal}$. 
% Covariates at the initial decision point were generated from an equicorrelated multivariate Gaussian distribution, while covariates at subsequent decision points were generated as functions of the previously observed covariates plus an independent error term. The propensity score was simulated from a logistic model. We considered both a linear outcome and a non-linear outcome with periodic behavior, as a function of covariates, prior actions and autoregressive (AR) errors. In certain simulation settings, where we considered a changepoint, we modified treatment to have a negative effect post changepoint, as opposed to a positive effect pre changepoint.

For a particular individual,

\textbf{Covariate generation.} $X^{'}_1 =(X_{11}^{'},...,X_{1p}^{'})^T$ at the first decision point is generated as an equicorrelated multivariate Gaussian vector with mean 0, Var$(X_{1j}^{'}) = 1$, Cov$(X_{1j}^{'}, X_{1j^{'}}^{'}) = \rho,~~ \forall ~ j \neq j^{'}$. The covariate vector $X_1 = (X_{11},...,X_{1p})^T$ is such that $X_{1j} = \Phi(X_{1j}^{'})$. At decision point $t>1, X_{t}^{'} = \boldsymbol\gamma X_{t-1} + \gamma_0\cdot A_{t-1}\cdot \mathbf{1} + \boldsymbol{\epsilon_1} + (A_{t-1} - \pi_{t-1})\boldsymbol{\epsilon_2}$, where $\boldsymbol{\gamma} = \text{diag}\{0.7, 0.7,..., 0.7  \}, \mathbf{1}$ is the unit vector, $\boldsymbol{\epsilon_1}$ and $\boldsymbol{\epsilon_2}$ are vectors of errors where $\epsilon_{1i} \sim \mathcal{N}(0,0.5)$, $\epsilon_{2i} \sim \mathcal{N}(0,0.25)$. We fix $\gamma_0 = 0.5$. Finally, $X_t = (X_{t1},...,X_{tp})^T$ is such that $X_{tj} = \Phi(X_{tj}^{'})$. We fix the number of covariates to be $P=50$ in  all scenarios explored.

\textbf{Propensity score.} We define a logit model for the propensity score at time t, $\pi_t(X_t, A_{t-1}) = \mathbb{P}(A_t =1 | X_t, A_{t-1}) = \frac{1}{1+e^{-(\boldsymbol{\beta}^T X_t + \beta_0 A_{t-1})}}$, where $\boldsymbol{\beta} = \{ 0.5, 0.3, 0, ..., 0 \}, \beta_0 = 0.25$. We set $A_0 = 0$. Finally, $A_t = \text{Ber}(\pi_t)$

\textbf{Outcome generation.} We define $Y_t$ as a function of $(A_t, A_{t-1}, X_t)$. For the case where we have a linear outcome, our definition is similar to past work \cite{boruvka2018assessing},

$Y_{t} = \theta_1 (A_{t-1} - \pi_{t-1}) + \theta_2^T X_t + (A_t - \pi_t)(\theta_3 + \theta_4^T X_t) + (A_t - \pi_t)\epsilon_{t,\text{trt}} + \epsilon_{t,Y_{t}}$, 
\\ where $\epsilon_{t,\text{trt}}$ and $\epsilon_{t,Y_{t}}$ are auto-regressive errors (AR(1)) with mean $0$ and variance $\sigma_y =0.05$. We set $\theta_1 = 0.5, \theta_3 = 0.7$, and define sparse effects for the covariates $\theta_2 = (2, 1, 0, 0, ..., 0), \theta_4 = (1, 2, 0, 0, ..., 0)$.

For the case when we have a non-linear outcome, we modify the linear outcome above as

$Y_t = Y_t +  \mathbb{I}\left( X_{t1} > 0.5 | X_{t2} > 0.5  \right) + |\text{sin}(\frac{t\pi}{7})|$

We define our non-linear outcome based on related work on ITEs \cite{lei2021, meta}.
% \cite{lei2021} and \cite{meta} have similar definitions for their non-linear outcome. 
We include some periodicity as well, as we are working in the longitudinal setting as opposed to the their cross-sectional setup.

\textbf{Changepoint outcome generation.} For the simulations with changepoints, we adopt a similar approach to Barber et al. (2023) \cite{BarberBE}, where before the changepoint $T \leq t_c$, we set $\theta_2 = (2, 1, 0, 0, ..., 0), \theta_4 = (1, 2, 0, 0, ..., 0)$, whereas after $T > t_c$, we modify $\theta_2 = (0, -2, -1, 0, 0, ..., 0), \theta_4 = (-1, -3, -2, 0, 0, ..., 0)$. This means that the action (treatment) suddenly has a negative impact post changepoint.

\newpage

\subsubsection{Quantile regression implementation}
\label{app:qr}

Unless otherwise specified, a linear model with an $L1$ penalty (LASSO) was used to predict the lower $(\alpha/2)$ and upper $(1-\alpha/2)$ weighted quantiles of the empirical distribution of calibration scores. As predictors, all covariates along with decision point were used. Signed errors $\hat{e}_{ij}$ of predicting the observed outcome were also leveraged with a lag of 3.

$\hat{e}_{ij} = \begin{cases}
    Y_{ij} - \hat{\mu}_1(X_{ij}, A_{ij}), ~~~ A_{ij} =1 \\
     Y_{ij} - \hat{\mu}_0(X_{ij}, A_{ij}), ~~~ A_{ij} =0
\end{cases}$

The \href{https://cran.r-project.org/web/packages/quantreg/index.html}{``quantreg"} package was used to implement quantile regression.

For neural network and random forest quantile regression implementations, please refer to Appendix \ref{app:nn} and \ref{app:rf} respectively.

\subsubsection{Neural network implementation}
\label{app:nn}

\textbf{Nuisance parameter estimation}.

As predictors, we use all covariates $(X_{ij})$, the current action $(A_{ij})$ and the previous action $A_{i,j-1}$. 

We use a neural network with a single hidden layer with number of input nodes equal to the number of predictors $(P)$ and number of nodes in the hidden layer equal to half of the the number of predictors in $X_{ij}$  $ ( = P/2)$. A kernel regularizer was used to account for sparsity in the data with an $L1$ penalty. The rectified linear unit (ReLU) activation function was used to compile the output and fed into a single node in the output layer with a linear activation function. We used the ``adam'' optimizer with a learning rate of 0.01. Mean squared error (MSE) was chosen to be the loss function. The training process was run using 100 epochs with a batch size of 32 and 20\% of the training data ($D_\varphi$) was reserved for validation.

\textbf{Quantile regression}.

The architecture remained the same as the case for nuisance parameter estimation. The only difference was in the choice of the loss function where we choose the ``pinball loss'' 
% as defined in 
\cite{cqr}. The predictor list remained the same as in Appendix \ref{app:qr}.

All neural network were implemented using the \href{https://cran.r-project.org/web/packages/keras3/index.html}{``keras3''} package. We note that the neural network predictions can be made better and more stable in performance by tuning the hyperparameters and running more epochs. However, since the purpose of the paper was to introduce a novel method and not to illustrate which machine learning method performs best, we leave this for future work.

\subsubsection{Random forest implementation}
\label{app:rf}

\textbf{Nuisance parameter estimation}.

As predictors, we use all covariates $(X_{ij})$, the current action $(A_{ij})$ and the previous action $A_{i,j-1}$. 

A random forest model with 500 trees and all other default settings was used to estimate the nuisance parameters. The R package \href{https://cran.r-project.org/web/packages/randomForest/index.html}{``randomForest''} was used to accomplish this step.

\textbf{Quantile regression}.

A random forest with 500 trees and all other default settings was used to estimate the quantiles. The R package \href{https://cran.r-project.org/web/packages/quantregForest/index.html}{``quantregForest''} was utilized.

\subsubsection{Compute resources}
\label{app:compute_resources}

Our method is not very computationally intensive. Every individual experiment can be run within $\approx$ 20 minutes. The linear models run faster and take $\approx$ 2-3 minutes, while the neural network and random forest models can take anywhere between 5-20 minutes to run, depending on the experiment and computing power. Due to the nature of the data which can involve thousands of observations and be up to 1-2 gigabytes (GB), we recommend using a computer system with at least 15-25GB Random-access Memory (RAM), to ensure fast processing time. One CPU is enough to run all results presented in the paper. Although not a necessity, a GPU can speed up neural network computations. We have run all results without GPU usage. All analyses were run using R version 4.2.2.

\newpage

% \newpage

\subsubsection{Data structure}
\label{app:sec_data_structure}

In this section, we provide a schematic representation of our assumed data structure along with the split of the training data. We visually show how the two types of predictions that our method is capable of producing are different. 

\begin{figure}[ht]
    \centering
    \includegraphics[width=\linewidth]{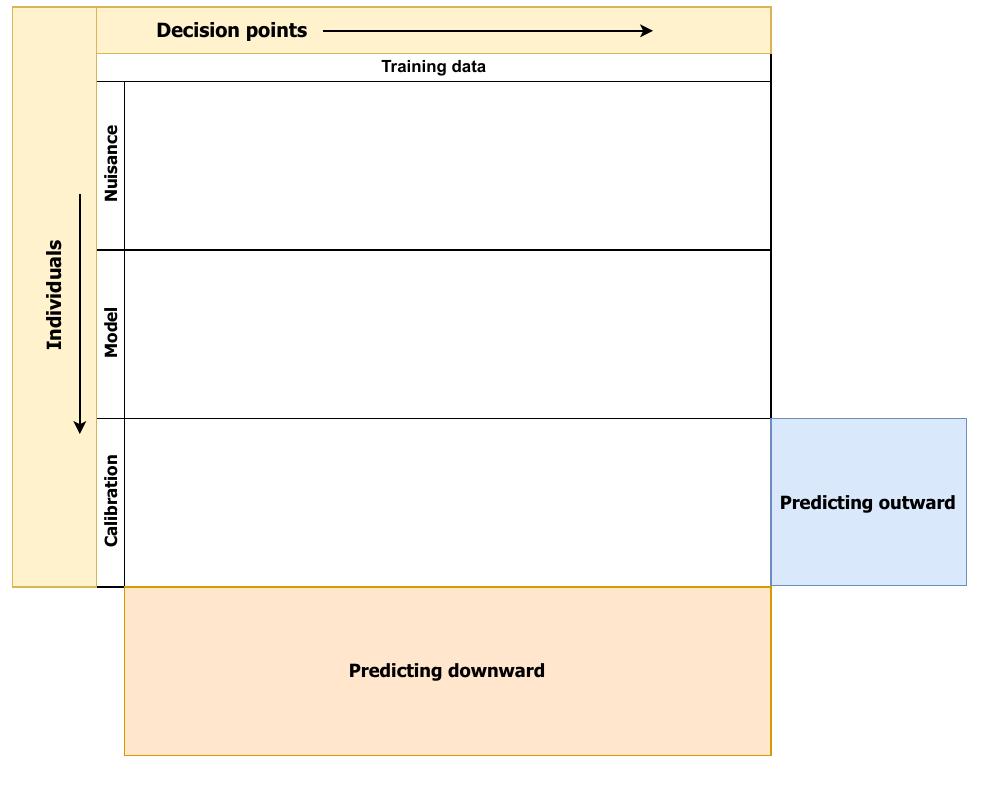}
    \caption{Illustration of our assumed data structure and distinction between the two types of predictions introduced in Section \ref{sec:sims}.}
    \label{fig:data_sructure}
\end{figure}

From Figure \ref{fig:data_sructure}, we can see how the training data are split into $D_{\varphi}$ (Nuisance), $D_\text{model}$ (Model) and $D_\text{cal}$ (Calibration) datasets. We always assume an equal split of the training data into the three mutually exclusive subsets and split by individuals (not by decision point). We see visually that predicting downward refers to providing prediction intervals for a new set of individuals at already observed decision points. Predicting outward corresponds to building intervals for an existing subset of individuals at future decision points.

We distinguish between the two types of predictions 

\textbf{Prediction intervals for ITE for a new set of individuals}. Suppose we use pilot study data as our training set and we are interested in creating prediction intervals for the ITE of a new set of individuals. We refer to this as ``predicting downward''.

\textbf{Prediction intervals for ITE at future decision points for an existing set of individuals}. Suppose we have observed a cohort of individuals for some time and our objective is to provide prediction intervals for the ITE for a subset of them at future decision points. We refer to this as ``predicting outward''.

\newpage

\subsubsection{Comparison of weighting schemes}
\label{app:weig_schemes}

\textbf{Description of weighting schemes.}

We have considered three different weighting schemes in addition to the equal weighting method -

\begin{enumerate}
    \item Equal weights (E) - Weights $(w_{ij})$ are assigned an equal weight.
    \item Decaying weights (D) - Weights $(w_{ij})$ are assigned weights that decay temporally with a factor $\psi$, $w_{ij} = \psi^{|t-j|}, ~\forall ~ i,j$, where $t$ is the decision point where we build the prediction interval.
    \item Decaying squared weights (DSQ) - Weights $(w_{ij})$ are assigned weights that decay faster temporally than D, $w_{ij} = \psi^{|t-j|^2}, ~\forall ~ i,j$, where $t$ is the decision point where we build the prediction interval.
    \item Decaying root weights (DRT) - Weights $(w_{ij})$ are assigned weights that decay temporally slower than D, $w_{ij} = \psi^{\sqrt{|t-j|}}, ~\forall ~ i,j$, where $t$ is the decision point where we build the prediction interval.
\end{enumerate}

After assignment of weights $(w_{ij})$, the weights are normalized $(\Tilde{w}_{ij})$. The weights are only dependent on the decision point and not the individual, hence it only depends on the subscript $j$.

\begin{figure}[h]
  \centering
  \includegraphics[width = 0.9\textwidth]{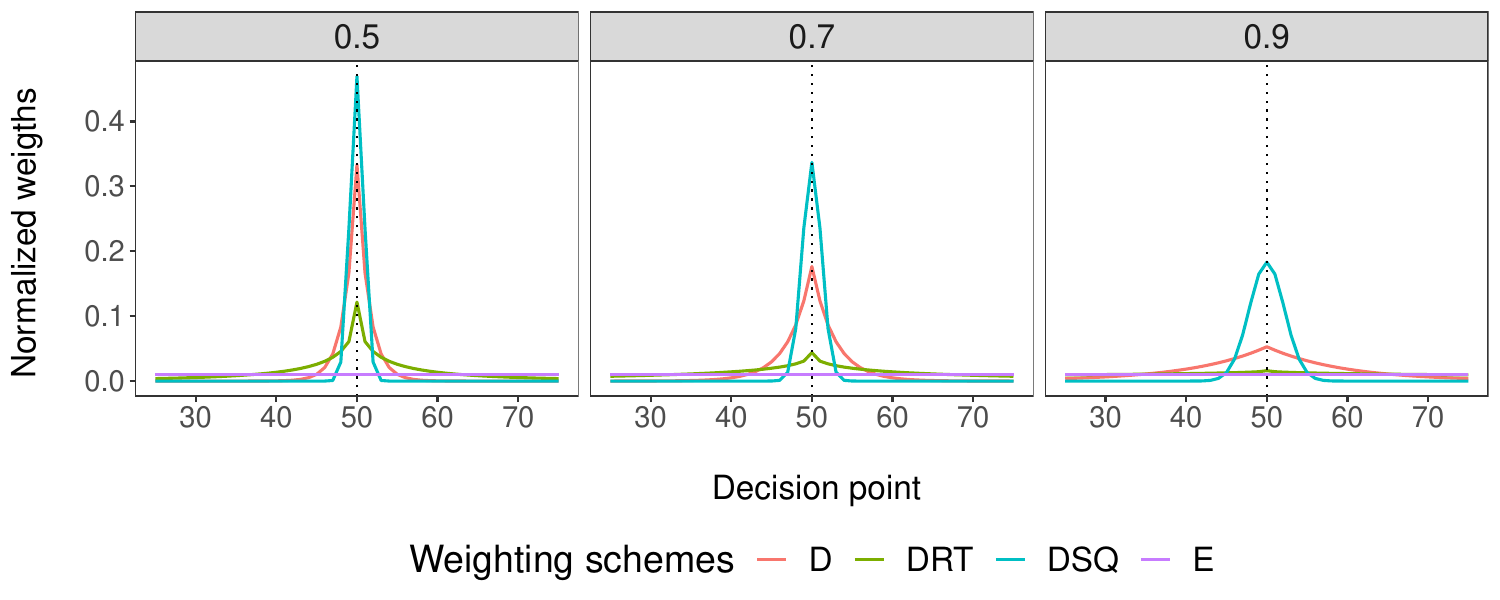}
  \caption{Normalized weights $(\Tilde{w}_{ij})$ assigned by the different weighting schemes for different values of $\psi = 0.5, 0.7, 0.9$, when creating prediction intervals for decision point 50 out of 100 decision points. \emph{Abbreviations:} E, Equal weights; D, Decaying weights; DSQ, Decaying squared weights; DRT, Decaying root weights.}
  \label{fig:diff_weig_schemes}
\end{figure}

From Figure \ref{fig:diff_weig_schemes}, we can see that there is a temporal decay of weights for all three weighting schemes - D, DRT, DSQ. This decay is only dependent on the distance of the decision points from the current decision point where the prediction interval is being built. The three methods of assigning weights are different in terms of the highest weights assigned, the rate of decay and the number of decision points they take into consideration. For instance, DSQ assigns a larger weight to decision points that are temporally closer to the decision point at which the interval is being created than D, but the weights decay at a much faster rate and it ends up assigning positive weights to a lower number of data points. On the contrary, DRT assigns a lower weight to decision points that are temporally closer to the decision point at which the interval is being created than D, but the weights decay at a much slower rate and it ends up assigning a positive weight to a higher number of data points.

\begin{figure}[ht]
  \centering
  
  \subfigure(a){\includegraphics[width =0.45\linewidth]{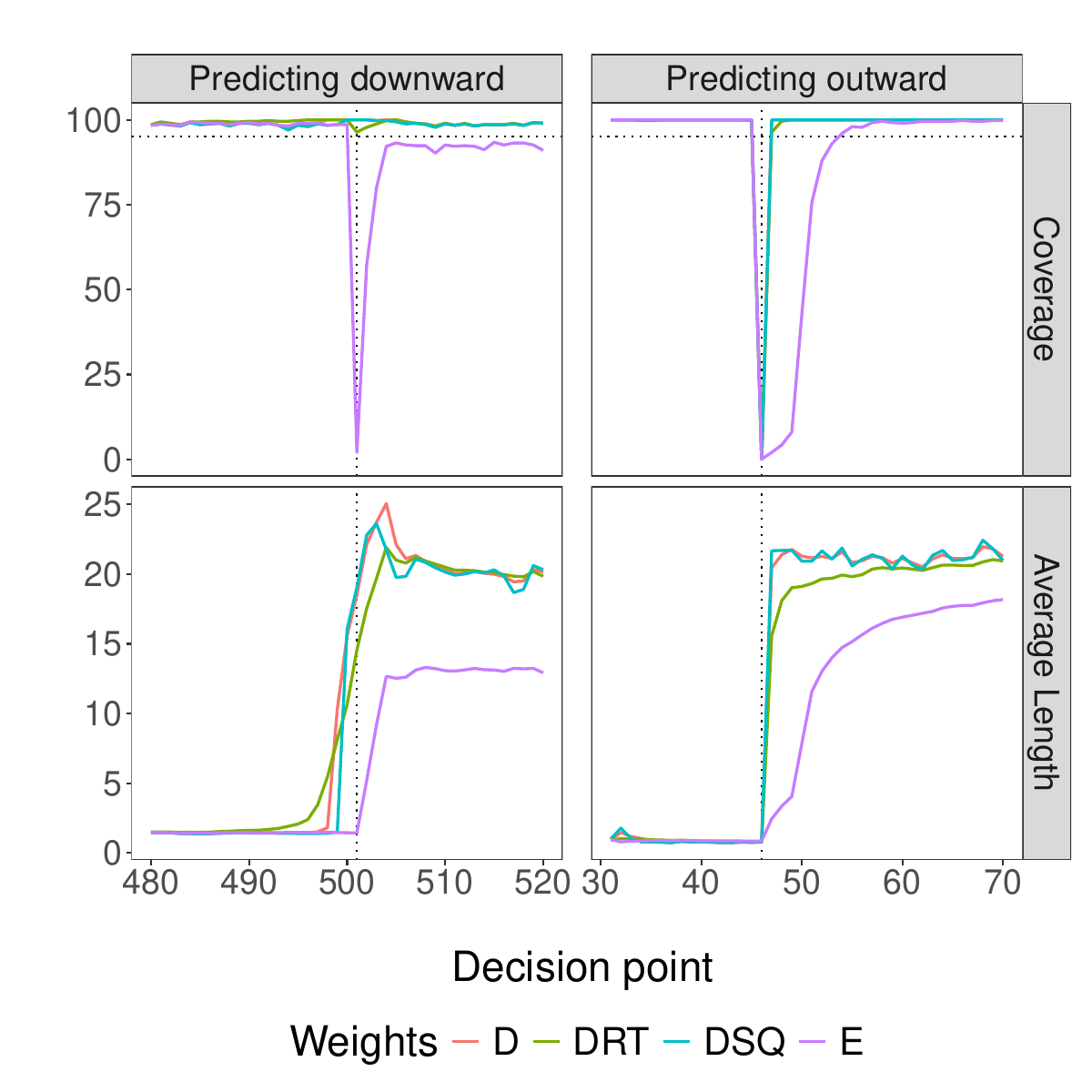}}
  \subfigure(b){\includegraphics[width =0.45\linewidth]{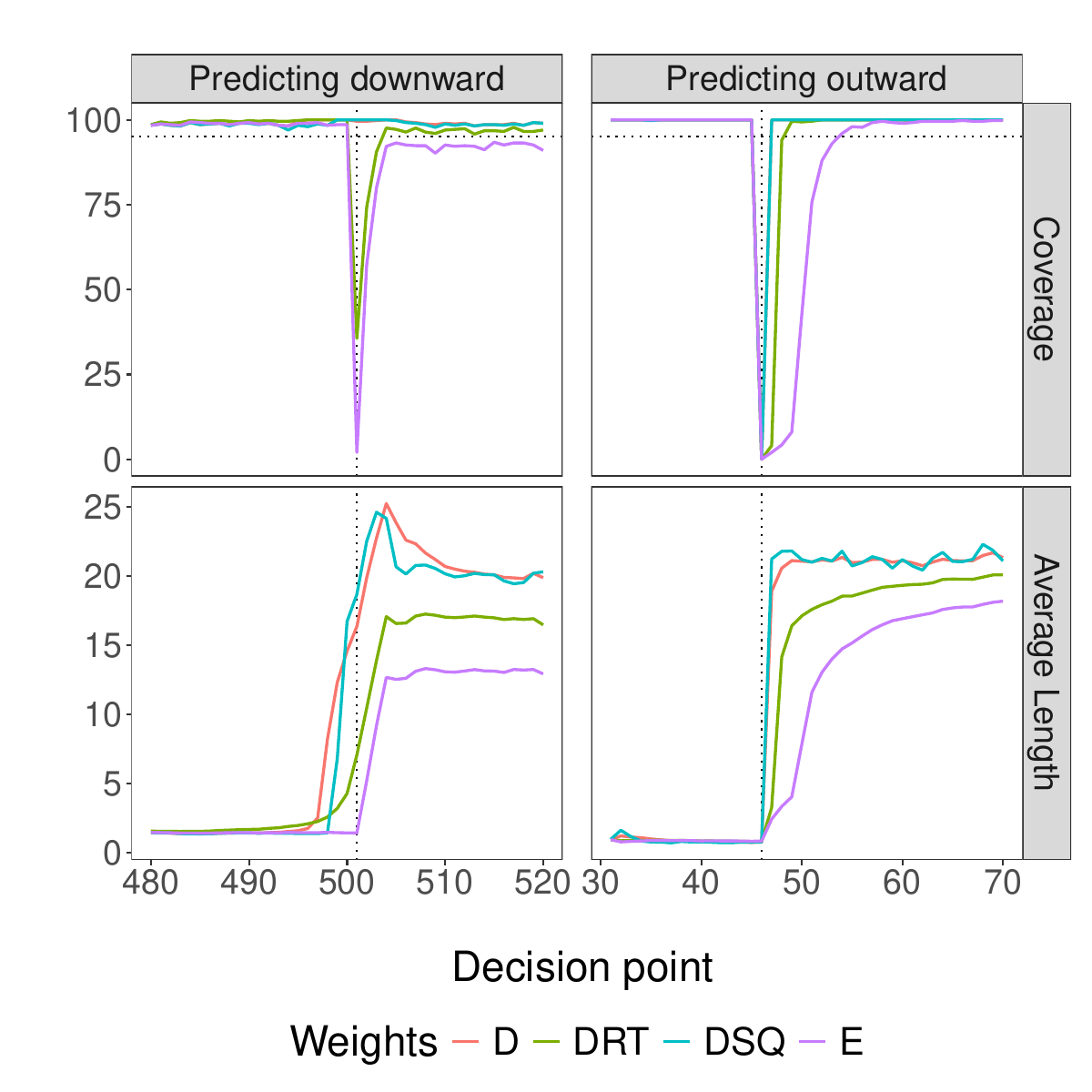}}
  \subfigure(c){\includegraphics[width =0.45\linewidth]{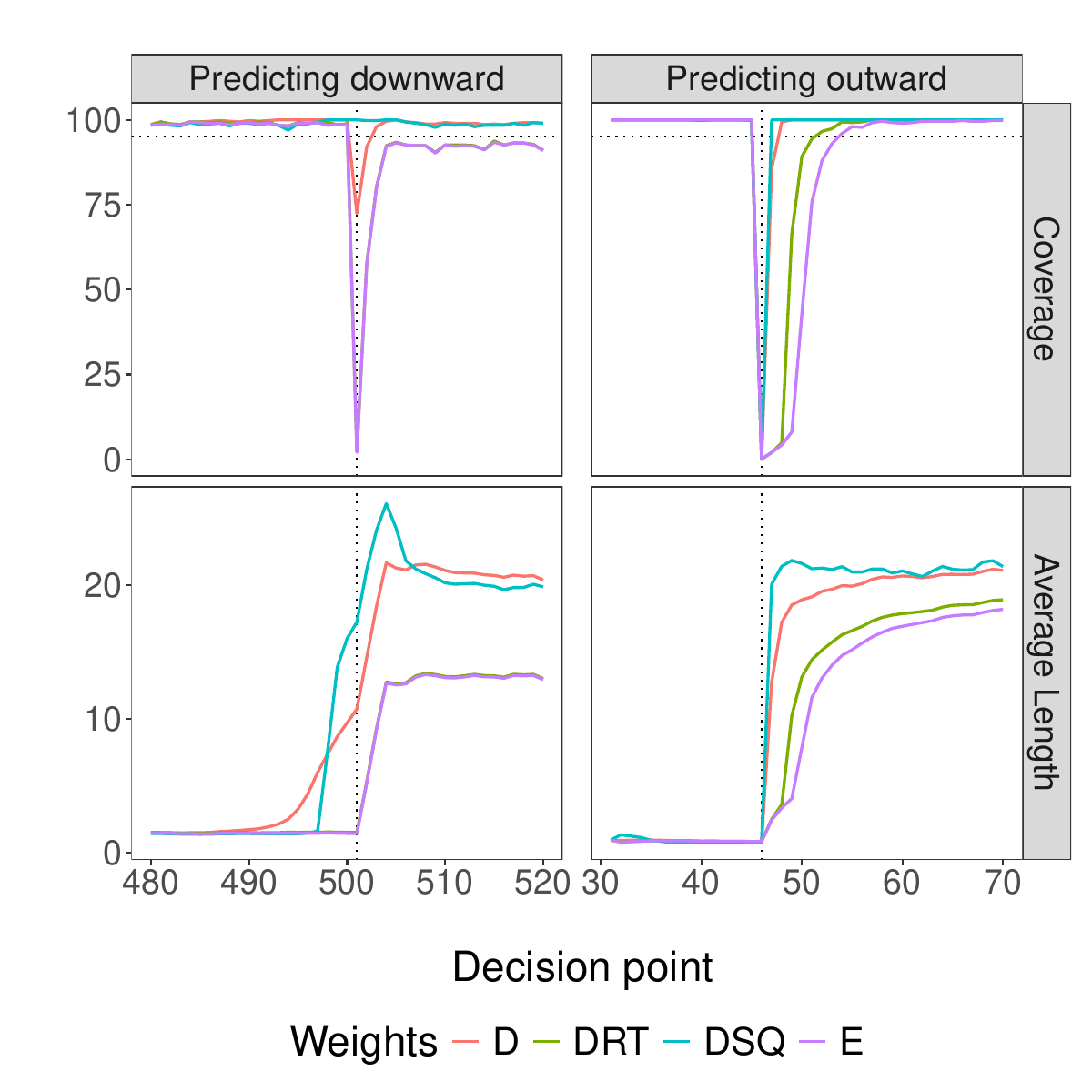}}
   
  \caption{Performance of different weighting schemes in the presence of a changepoint, where the simulation settings are the same as described in Section \ref{sec:sim_chgpts} for different values of $\psi= 0.5, 0.7, 0.9$. Horizontal dotted lines represent target coverage and vertical dotted lines represent the changepoints. Refer to Appendix \ref{app:sec_data_structure} for details about the two types of predictions. \emph{Abbreviations:} E, Equal weights; D, Decaying weights; DSQ, Decaying squared weights; DRT, Decaying root weights.
  \textbf{(a)} $\psi = 0.5$. 
  \textbf{(b)} $\psi = 0.7$.
  \textbf{(c)} $\psi = 0.9$}.
  \label{fig:psi_changepoint}
\end{figure}

\textbf{Comparison of performance of weighting schemes.}

In the settings described in Section \ref{sec:sim_chgpts}, where we have a changepoint, we have evaluated performance of the linear model with different weighting strategies. For completeness, we describe the settings again.

We evaluate two instances in our linear outcome setup -

\textbf{Predicting outward}. We generate data on $T=90$ decision points but only use data until $T=30$ to train the models. We define the changepoint at $t_c=45$. In this case, we do not observe a changepoint in our training data, and this can be considered to be a case of model misspecification. As our test dataset, we use $D_\text{cal}$ with all decision points post 30. 

\textbf{Predicting downward}. We generate data on $T=520$ decision points and the algorithm is trained on all decision points in $D_\text{tr}$. We define the changepoint at $t_c=500$. In this case, we do observe a changepoint in our training data, but the data prior to changepoint constitutes more than 95\% of the training data. We use $D_\text{test}$ as our test dataset. 

Since our objective is getting closer to the desired coverage level, we observe that across all values of $\psi$ considered, a decaying weights strategy (D, DSQ, DRT) offers performance superior to that of the equal weights. E performs poorly in such scenarios where a changepoint is present - very common in MRT settings - motivating the need for a more informed weighting strategy. E is slower to react to the change point when predicting outward and does not achieve target coverage levels after changepoint when predicting downward. In general, D and DSQ perform best, followed by DRT.  Based on these findings, we recommend using D or DSQ with $\psi=0.7$, as it is robust to model misspecification and changes in the data generating process. For all results reported in the main text after Section \ref{sec:sim_chgpts}, we have used D with $\psi=0.7$.

\newpage

\subsubsection{Additional results}
\label{app:add_results}

Results provided in Figure \ref{fig:weights} have been presented in tabular format in Table \ref{tab:weight_preds}. We are interested in observing differences between two different weighting schemes in the presence of changepoints. Here, the results presented have been aggregated across all individuals and decision points in our test dataset.

\begin{figure}[ht]
    \centering
    \includegraphics[width=0.75\linewidth]{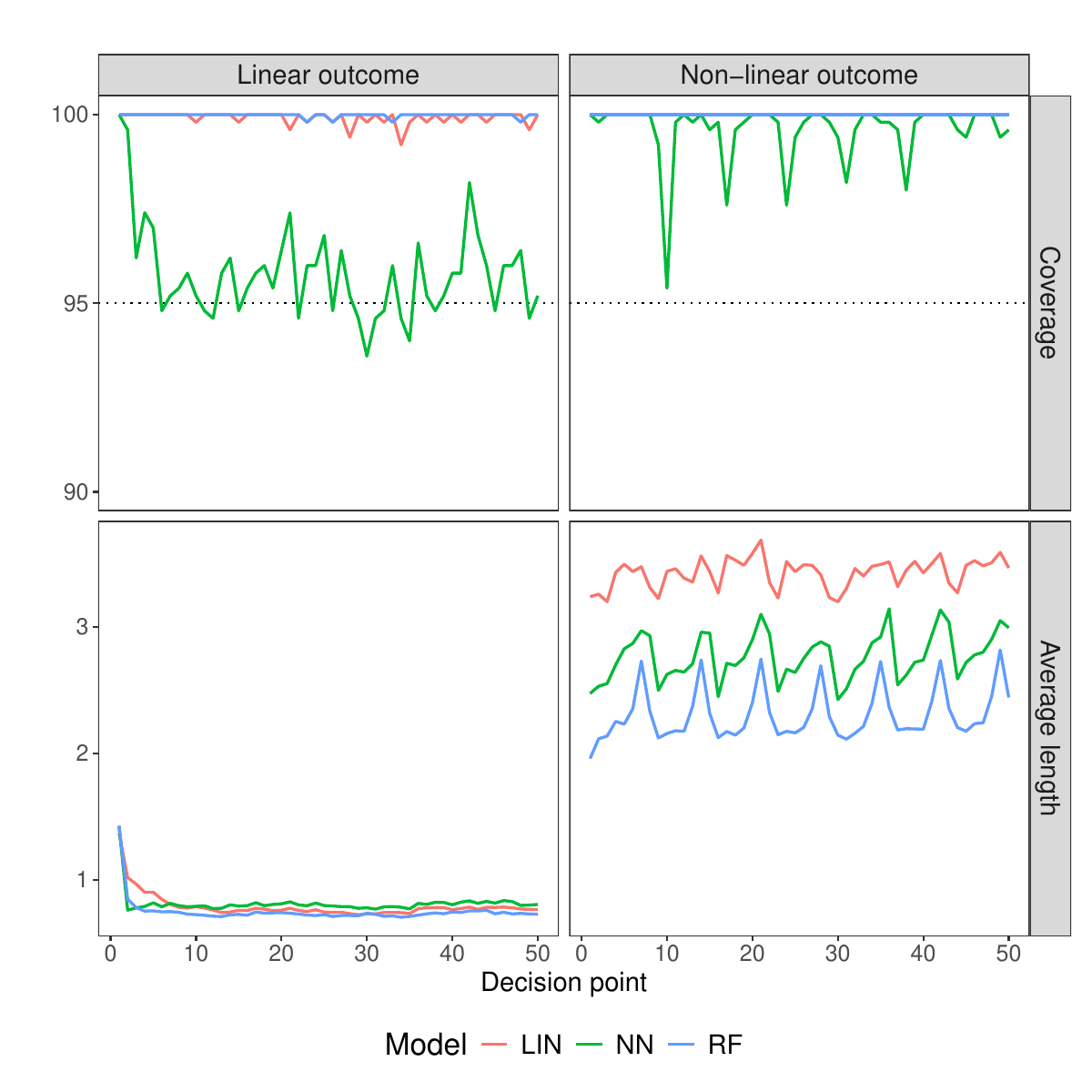}
    \caption{Diagrammatic representation of how coverage and lengths differ over decision point for results corresponding to Table \ref{tab:ml_preddown}, where the Performance of ML models are evaluated on a linear and non-linear outcome. Results are for predicting downward and are averaged over $N=500$ individuals in our simulated test dataset at every decision point. Horizontal lines represent target coverage. \emph{Abbreviations:} LIN, Linear model with $L1$ penalty (LASSO); NN, Neural network; RF, Random forest.}
    \label{fig:ml_preddown}
\end{figure}

We provide a diagrammatic representation in Figure \ref{fig:ml_preddown} of competing ML models as seen in Table \ref{tab:ml_preddown}. Here, we evaluate the Performance of ML models in handling a non-linear outcome. Results have been averaged over 500 individuals at every decision point. 

\begin{table}[ht]
\caption{Results corresponding to Figure \ref{fig:weights}, where we compare algorithm performance when using two different weighting schemes in the calibration step in the presence of changepoints. All results presented in this table have been averaged over all individuals and over all decision points in the test dataset. \emph{Abbreviations}: Cov, Coverage; PCov, Pseudo-outcome coverage; AL, Average length. \\}
\centering{
\small{
\begin{tabular}{@{}cllllll@{}}
\toprule
\multirow{3}{*}{}     & \multicolumn{6}{c}{Prediction type}                        \\ \cmidrule(l){2-7} 
                         Weights    & \multicolumn{3}{c}{Downward} & \multicolumn{3}{c}{Outward} \\ \cmidrule{2-7}
                             & Cov      & PCov     & AL     & Cov     & PCov    & AL      \\ \cmidrule(r){1-7}
\multicolumn{1}{l}{Equal}    & 98.58    & 95.34    & 1.85   & 91.50   & 81.81   & 12.06   \\
\multicolumn{1}{l}{Decaying} & 98.97    & 95.45    & 2.22   & 98.33   & 93.26   & 15.68   \\ \bottomrule 
\end{tabular}
}}
\label{tab:weight_preds}
% \end{centering}
\end{table}

\newpage

\subsubsection{Sensitivity analysis of interval lengths and coverage}
\label{app:sensitivity_analysis}

We conduct sensitivity analysis to observe how coverage and average length are affected when we vary errors related to our outcome and the size of the training.

\textbf{Different variances of the error associated with the outcome}

Recall that we had defined our outcome to have auto-regressive errors with $\epsilon_{t,Y_t}$ and $\epsilon_{t,trt}$, which are two independent autoregressive (AR(1)) white noise processes with mean $0$ and variance $\sigma_y = 0.05$. In this section, we evaluate empirical differences in interval length and coverage for other choices of the variance. We consider the case where we have a linear outcome and our method uses a linear model with an $L1$ penalty (LASSO) for estimation of nuisance functions as well as quantile regression. All other simulation settings remain unchanged and we generate data for $N=1500$ individuals in $D_\text{tr}$, over 50 decision points for predicting downward (and 100 decision points for predicting outward). For predicting outward, models are fit on the first 50 decision points and predictions are made for the last 50 decision points of $D_\text{cal}$, which constitutes our test dataset in this case. Results in Table \ref{tab:sensitivity_sigma} are averaged over 500 individuals in the test dataset and over 50 decision points. First, we note that in all scenarios, our coverage guarantees hold. Also, we observe that the average lengths of our prediction intervals increases with increase in the variance associated with the outcomes, which is in line with our intuition.

\begin{table}[ht]
\caption{\small{Results corresponding to different choices for the variance of the error $(\sigma_y)$ associated with a linear outcome, when a linear model is used throughout in our method. All other simulation settings remain unaltered and results are averaged over 500 individuals and 50 decision points in test data. \emph{Abbreviations}: Cov, Coverage; PCov, Pseudo-outcome coverage; AL, Average length.} \\}
\centering{
\small{
\begin{tabular}{@{}cllllll@{}}
\toprule
\multirow{3}{*}{}     & \multicolumn{6}{c}{Prediction type}                        \\ \cmidrule(l){2-7} 
                         $\sigma_y$    & \multicolumn{3}{c}{Downward} & \multicolumn{3}{c}{Outward} \\ \cmidrule{2-7}
                             & Cov      & PCov     & AL     & Cov     & PCov    & AL      \\ \cmidrule(r){1-7}
\multicolumn{1}{l}{0.05}    & 99.91    & 95.09    & 0.79   & 99.92   & 94.51   & 0.81   \\
\multicolumn{1}{l}{0.1}  & 99.87    & 95.06    & 1.32   & 99.87   & 94.74   & 1.33 \\
\multicolumn{1}{l}{0.25}    & 99.75    & 95.06    & 3.12   & 99.68   & 94.92   & 3.12   \\
\multicolumn{1}{l}{0.5} & 99.73    & 95.10    & 6.20   & 99.63   & 94.92   & 6.16 \\ \bottomrule
\end{tabular}
}}
\label{tab:sensitivity_sigma}
% \end{centering}
\end{table}

\textbf{Different sample sizes for training data $D_\text{tr}$}

In this section, we are going to look at the effect of different sample sizes for our training dataset $D_\text{tr}$ on model performance when we consider a non-linear outcome. We are going to focus on the predicting downward case and keep all other simulation settings unaltered. We generate data on $N$ individuals for 50 decision points. Results in Table \ref{tab:sensitivity_n} have been averaged over all 500 individuals in our test dataset and over all 50 decision points. We first note that irrespective of the size of the training set, we are able to achieve our desired coverage level. We also notice a trend towards getting tighter intervals as the sample size increases especially for the more sophisticated machine learning models like neral networks and random forests. This is similar to the findings in the literature \cite{kivaranovic2020conformal}.

\begin{table}[ht]
\caption{\small{Results corresponding to different choices for the size of the training dataset $D_\text{tr}$, when looking at the non-linear outcome. All other simulation settings remain unaltered and results are averaged over 500 individuals in the test dataset and 50 decision points. \emph{Abbreviations}: Cov, Coverage; PCov, Pseudo-outcome coverage; AL, Average length; LIN, Linear model with $L1$ penalty (LASSO); NN, Neural network; RF, Random forest.}\\}
\centering{
\small{
\begin{tabular}{@{}cllllllllllll@{}}
\toprule
\multirow{3}{*}{}     & \multicolumn{12}{c}{N}                        \\ \cmidrule(l){2-13} 
                         Model    & \multicolumn{3}{l}{60} & \multicolumn{3}{l}{300} & \multicolumn{3}{l}{600} & \multicolumn{3}{l}{1500} \\ \cmidrule{2-13}
                             & Cov      & PCov     & AL     & Cov     & PCov    & AL & Cov      & PCov     & AL     & Cov     & PCov    & AL      \\ \cmidrule(r){1-13}
\multicolumn{1}{l}{LIN}    & 100.00    & 95.99    & 3.77   & 100.00   & 95.42   & 3.47 & 100.00      & 95.42     & 3.44     & 100.00     & 95.50    & 3.43  \\
\multicolumn{1}{l}{NN}  & 99.66    & 94.66    & 3.23   & 98.32   & 95.09   & 3.11 & 97.90      & 95.12     & 2.88     & 98.96     & 95.15    & 2.61\\
\multicolumn{1}{l}{RF}    & 100.00    & 94.76    & 3.13   & 100.00   & 95.29   & 2.63 & 100.00      & 95.31     & 2.48     & 100.00     & 95.18    & 2.28  \\ \bottomrule
\end{tabular}
}}
\label{tab:sensitivity_n}
% \end{centering}
\end{table}

\newpage 

\subsection{Real data example: IHS (2018)}
\label{app:ihs}

\subsubsection{Data description}
\label{app:ihs_desc}

\begin{figure}[ht]
    \centering
    \subfigure[]{\includegraphics[height=1.4in,width=0.4\textwidth]{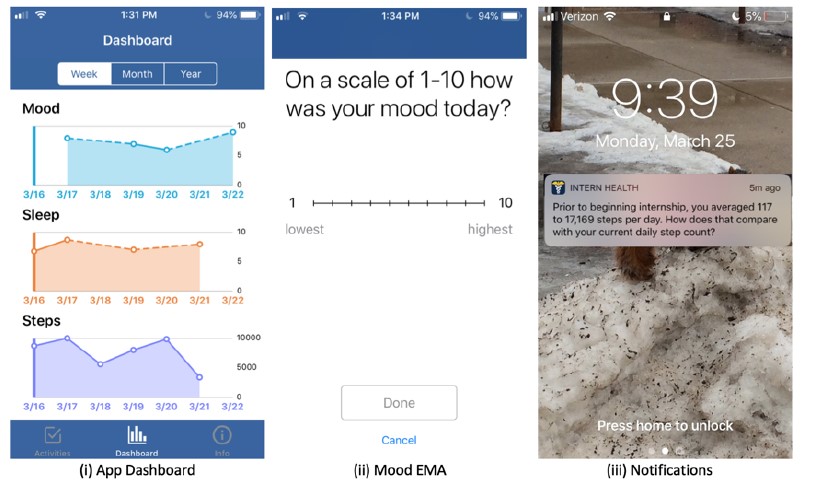}} 
    \subfigure[]{\includegraphics[height=1.2in,width=0.45\textwidth]{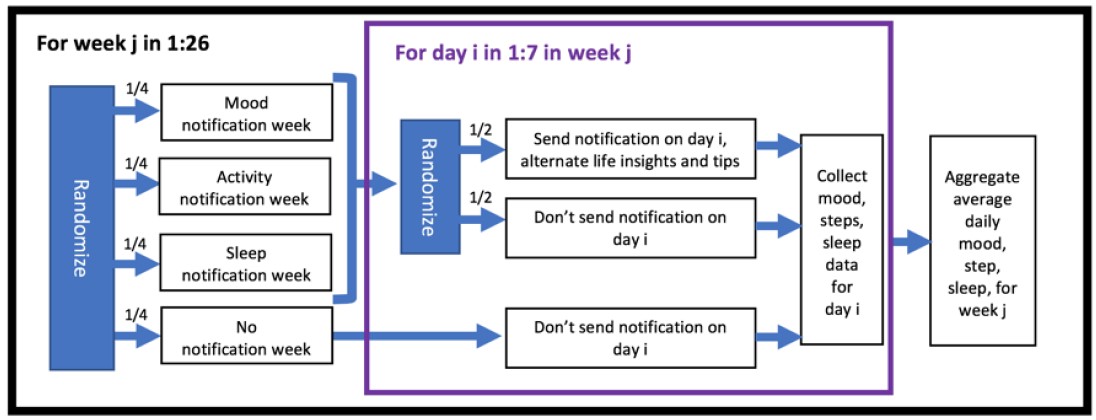}} 
    \caption{The IHS 2018 setup. (a) The app and notification screens as seen by participants. (b) The treatment assignment and data collection mechanism.}
    \label{fig:ihs_app_rand}
\end{figure}

The Intern Health Study (IHS) 2018 is an MRT, where 1563 medical interns were enrolled and data on sleep and activity was collected daily over a period of 6 months from June to December, on a ``Fitbit Charge 2'' device \cite{necamp2020assessing}. Interns were randomized to either of the three treatments (actions) or no treatment at a weekly level. All three actions or no action were equally likely to be assigned to an intern with probability $1/4$. Daily data were then aggregated at a weekly level. Figure \ref{fig:ihs_app_rand}\footnote{Figures taken from NeCamp et al. (2020) \cite{necamp2020assessing}} gives a schematic representation of how treatment assignments and data collection transpired.

Mood was measured on a scale of 1-10, where 1 referred to the worst and 10 the best possible mood for an intern. Step count was measured in '1000s and sleep was measured as the average daily hours of sleep. The two outcomes - step count and sleep, were transformed as follows

\begin{align}
    \text{Step count} = \sqrt[3]{\text{Step count}} \nonumber \\
\text{Sleep} = \sqrt{\text{Sleep (minutes)}}
\end{align}

% \footnotetext{hahaha}

\begin{figure}[ht]
    \centering
    \includegraphics[width=0.5\linewidth]{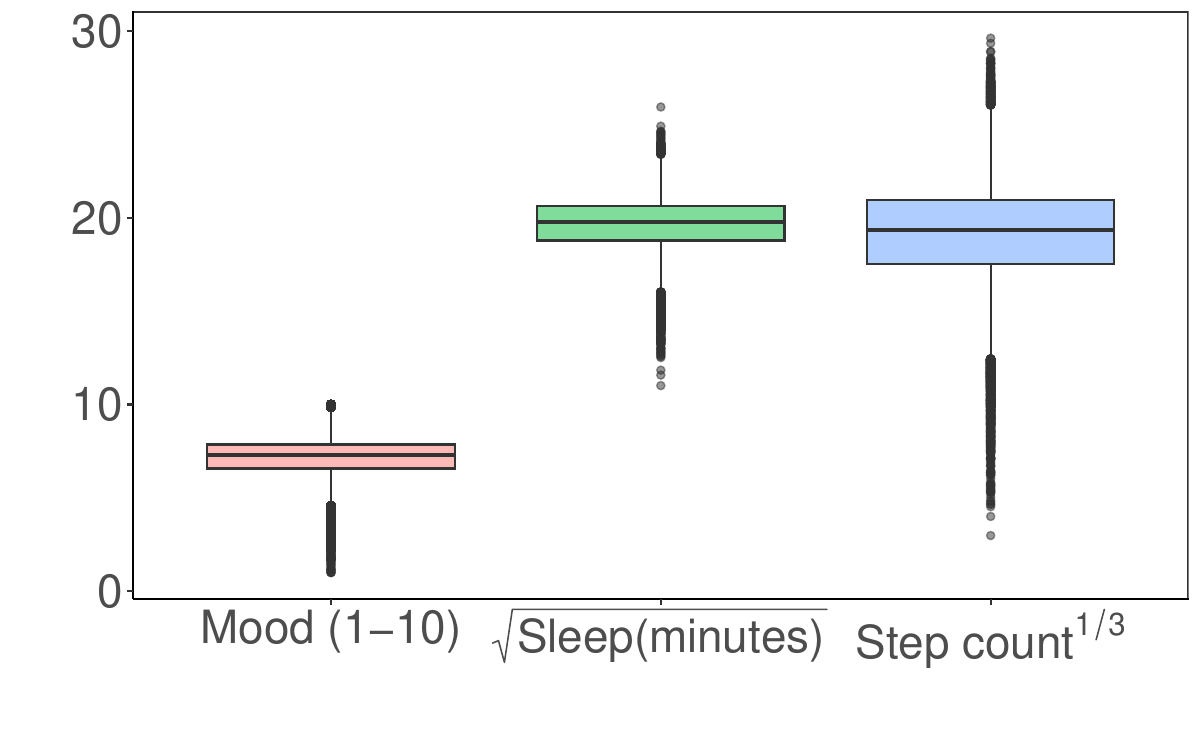}
    \caption{Boxplots for the transformed outcomes in IHS 2018, after single imputation. Mood was measured on a scale of 1-10, sleep was measured as $\sqrt{\text{Sleep (minutes)}}$, and step count was measured as $\sqrt[3]{\text{Step count}}$. All outcomes were aggregated at a weekly level.}
    \label{fig:ihs_boxplot}
\end{figure}

To tackle the problem of missing data in IHS 2018, we have used single imputation.
% to account for the missingness.
Boxplots for the transformed outcomes can be found in Figure \ref{fig:ihs_boxplot}.

% \begin{figure}[ht]
%     \centering
%     \includegraphics[width=0.5\linewidth]{Images/ihs_boxplot.pdf}
%     \caption{Boxplots for the transformed outcomes in IHS after single imputation.}
%     \label{fig:ihs_boxplot}
% \end{figure}

For more details about IHS 2018, we refer the reader to NeCamp et al. (2020) \cite{necamp2020assessing}.

\subsubsection{Results}

As predictors, we used biological sex, previous week's outcomes, and pre-internship mood, sleep and step count levels for both nuisance parameter estimation as well as quantile prediction.

We split the 1563 interns into training ($D_\text{tr}$) and test ($D_\text{test}$) data splits with 75\% and 25\% of the interns respectively in each dataset. We split $D_\text{tr}$ equally into thirds to form $D_\varphi, D_\text{model}$ and $D_\text{cal}$.

% Coverage guarantees provided are for our created pseudo-outcomes, which are different for the different ML methods used. 
Coverage is evaluated on our created pseudo-outcomes (which are model-specific). As we have seen in Section \ref{sec:pseudo_worstcase}, coverage of the true ITE is always greater than pseudo-outcome coverage. We leverage this idea and test the performance of our method on pseudo-outcome coverage as a surrogate for real ITE coverage, as real ITE is never observed in practice.

\textbf{Predicting downward}. After training the model using $D_\text{tr}$, we get predictions for $D_\text{test}$ at all decision points. Results have been averaged over all interns and decision points in $D_\text{test}$ and are provided in Table \ref{tab:ihs_preddown}. We see that overall, the linear model seems to be doing quite well and provides performance competitive to that of the neural network. The random forest displays poorer performance likely due to the fact that we are working with a very small predictor pool. 

\textbf{Predicting outward}. Here, we only use the first 16 weeks worth of data in $D_\text{tr}$ to train our algorithm. We expand $D_\text{cal}$ to contain information from $D_\text{test}$ for all weeks prior to week 16 and similarly expand $D_\text{test}$ to include information on interns from $D_\text{cal}$ post week 16. The results have been averaged over all interns and decision points in $D_\text{test}$ (10 weeks worth of data; week 17-26) and are provided in Table \ref{tab:ihs_predout}. Our observations are similar to the predicting downwards case, where the linear model seems to have better or comparable performance with the other machine learning models.

As a reminder, performance is evaluated using average length and pseudo-outcome coverage, which we are considering a surrogate for actual ITE coverage.

\begin{table}[ht]
\caption{Performance of ML models when creating prediction intervals for IHS 2018 data when predicting outward. Results averaged over all individuals and decision points (weeks) in the test data. Best performing model's average length bolded. \emph{Abbreviations}: PCov, Pseudo-outcome coverage; AL, Average length; LIN, linear model (LASSO); NN, Neural network; RF, Random forest.\\}
\centering{
\begin{tabular}{@{}lllllll@{}}
\toprule
\multirow{2}{*}{} & \multicolumn{6}{c}{Outcome}                                                      \\ \cmidrule(l){2-7} 
                    Model   & \multicolumn{2}{l}{Step count} & \multicolumn{2}{l}{Sleep} & \multicolumn{2}{l}{Mood} \\ \cmidrule(r){2-7}
                       & PCov         & AL          & PCov         & AL          & PCov         & AL         \\
                       \cmidrule(r){1-7}
LIN                    & 94.76       & \textbf{17.43}       & 94.71       & \textbf{10.38}       & 94.65       & 5.72       \\
NN                     & 95.11       & 18.35       & 94.93       & 10.94       & 94.39       & \textbf{5.70}       \\
RF                     & 94.97       & 18.31       & 94.97       & 10.72       & 94.66       & 5.97       \\ \bottomrule \\
\end{tabular}
}
\label{tab:ihs_predout}
% \end{centering}
\end{table}

\newpage

\subsection{Intern Health Study (IHS), 2020}
\label{app:ihscomp}

\subsubsection{Summary}

The Intern Health Study (IHS) 2020 is an MRT, where sleep and activity data were collected on 1936 interns (who were grouped into 191 teams) from April to June in 2020. Data were aggregated at the weekly level for each team, which was considered to be the unit of analysis \cite{wang2023effectiveness}. Each team was randomized to competition or no competition every week (with equal probability 0.5) and the effectiveness of team competition on step count and sleep (minutes) was assessed. We build prediction intervals for ITEs of both outcomes separately and consider the effect of team competition vs no team competition on the outcomes.

\begin{table}[ht]
\caption{Performance of ML models when creating prediction intervals for IHS (2020) data when predicting downward. Results averaged over all individuals and decision points in the test data. Best performing model's average length bolded. \emph{Abbreviations}: PCov, Pseudo-outcome coverage; AL, Average length; LIN, Linear model with $L1$ penalty (LASSO); NN, Neural network; RF, Random forest.\\}
\centering
{\small
\begin{tabular}{@{}lllllll@{}}
\toprule
\multirow{2}{*}{} & \multicolumn{4}{c}{Outcome}                                                      \\ \cmidrule(l){2-5} 
                    Model   & \multicolumn{2}{l}{Step count} & \multicolumn{2}{l}{Sleep}  \\ \cmidrule(r){2-5}
                       & PCov         & AL          & PCov         & AL             \\
                       \cmidrule(r){1-5}
LIN                    & 94.26       & \textbf{4842}.54       & 94.91       & \textbf{139.46}             \\
NN                     & 93.42       & 5138.29       & 95.14       & 178.21         \\
RF                     & 94.22       & 4991.42       & 94.97       & 151.06       \\ \bottomrule 
\end{tabular}
}

\label{tab:ihscomp_preddown}
% \end{centering}
\end{table}

Table \ref{tab:ihscomp_preddown} shows performance of different machine learning algorithms on a hold out test data subset (roughly 25\%) from IHS 2020, when predicting downward (averaged over all individuals and decision points/weeks). Results for predicting outward can be found in Appendix \ref{app:ihscomp_res}. Again, coverage is estimated based on the pseudo-outcome coverage, where pseudo-outcomes were created by the different models, since we never observe true ITE in reality. For a more detailed discussion of results, please see Appendix \ref{app:ihscomp_res}.

% \subsection{Real data example: IHS (2020)}
% \label{app:ihscomp}

\subsubsection{Data description}

The Intern Health Study (IHS) 2020 is an MRT, where a total of 1936 medical interns were grouped into 191 teams \cite{wang2023effectiveness}. Data were collected on step count and sleep related data from April to June 2020. Each team (the unit of analysis) was randomly assigned to be in competition against another team (treatment) vs not in competition for a particular week, with equal probability 1/2. Teams in competition in a particular week would compete in step count or sleep (minutes), which was randomly assigned with probability 1/2. The effectiveness of treatment (competition) was evaluated on average weekly data for a particular team. 

\begin{figure}[ht]
    \centering
    \includegraphics[width=0.75\linewidth]{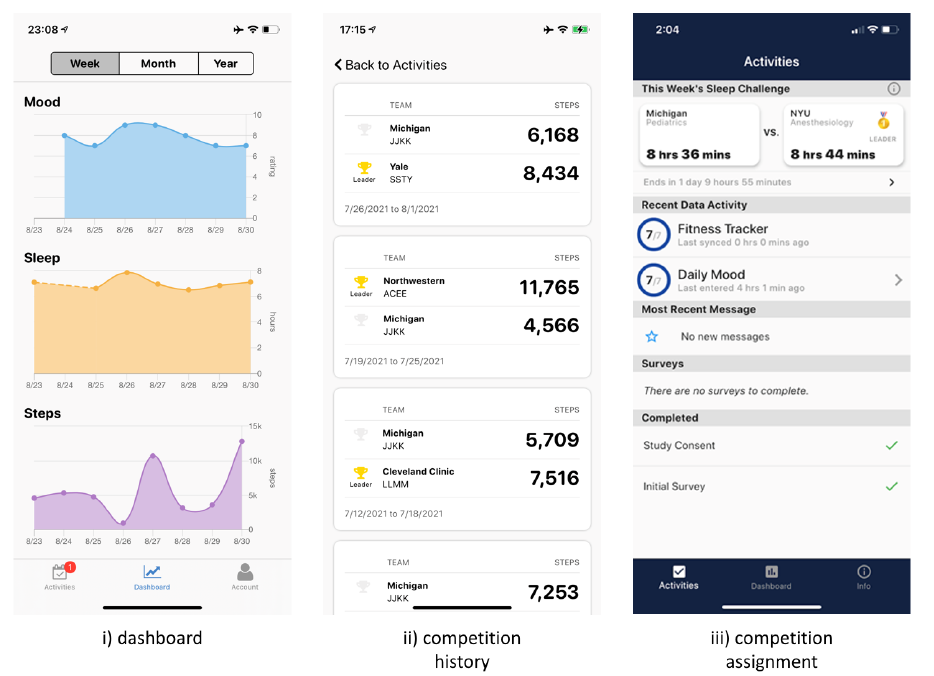}
    \caption{Screenshots of the app's dashboard, competition history and competition assignment.}
    \label{fig:ihs_dashcomp}
\end{figure}

Step count was measured in '1000s and sleep was measured as average daily minutes of sleep. Mood was measured on a scale of 1-10, with 1 being the worst and 10 being the best possible mood of a participant. Unlike IHS 2018, there were no mood related treatments, so we do not build prediction intervals for mood in this case. Transformations were not used to be consistent with the literature \cite{wang2023effectiveness}. Figure \ref{fig:ihs_dashcomp}\footnote{Figures taken from Wang et al. (2023) \cite{wang2023effectiveness}} provides a screenshot of the screen seen by participants when using the app through which data were collected. 

\begin{figure}
    \centering
    \includegraphics[width=0.5\linewidth]{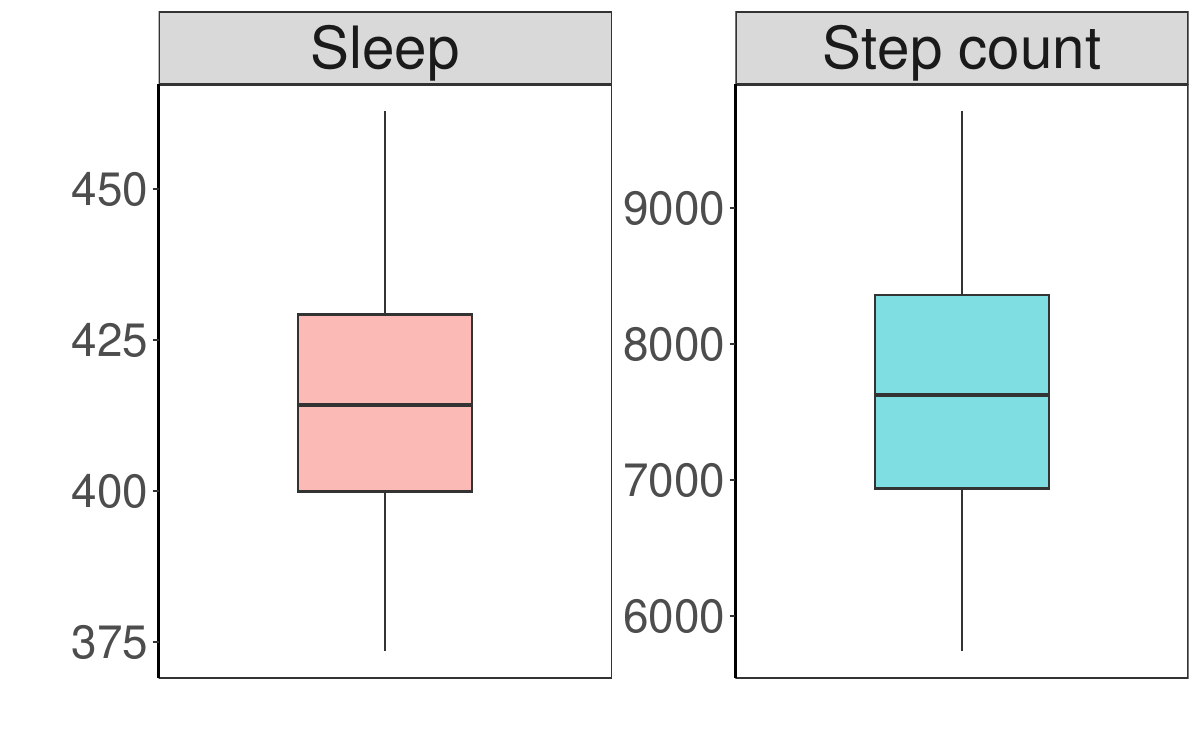}
    \caption{Boxplots for the outcomes in IHS 2020 after single imputation and outlier handling. Sleep was measured in minutes and step count was measured in '1000s. All outcomes were aggregated at the weekly team level.}
    \label{fig:ihscomp_boxplot}
\end{figure}

Missing data were handled using single imputation. We removed the bottom and top 5\% of the data w.r.t each outcome to handle outliers. Boxplots of the two outcomes can be found in Figure \ref{fig:ihscomp_boxplot}. 

For more details about IHS 2020, we refer the reader to Wang et al. (2023) \cite{wang2023effectiveness}.

\subsubsection{Results}
\label{app:ihscomp_res}

As predictors, we used previous week's outcomes on mood, sleep and step count for both nuisance parameter estimation as well as quantile prediction. The aggregation of teams at the weekly level made it difficult to engineer other relevant features that would be important for prediction.

We split the 191 teams into training ($D_\text{tr}$) and test ($D_\text{test}$) data splits with 75\% and 25\% of the teams respectively in each dataset. We split $D_\text{tr}$ equally into thirds to form $D_\varphi, D_\text{model}$ and $D_\text{cal}$.

% Coverage guarantees provided are for our created pseudo-outcomes, which are different for the different ML methods used. 
Coverage is evaluated on our created pseudo-outcomes (which are model-specific). As we have seen in Section \ref{sec:pseudo_worstcase}, coverage of the true ITE is always greater than pseudo-outcome coverage. We leverage this idea and test the performance of our method on pseudo-outcome coverage as a surrogate for real ITE coverage, as real ITE is never observed in practice.

\textbf{Predicting downward}. After training the model using $D_\text{tr}$, we get predictions for $D_\text{test}$ at all decision points. Results have been averaged over all interns and decision points in $D_\text{test}$ and are provided in Table \ref{tab:ihscomp_preddown}. We see that overall, the linear model seems to be doing better in terms of coverage and average length. The random forest and neural network models display poorer performance in comparison, likely due to the fact that we are working with a very small predictor pool. 

\textbf{Predicting outward}. Here, we only use the first 8 weeks worth of data in $D_\text{tr}$ to train our algorithm. We expand $D_\text{cal}$ to contain information from $D_\text{test}$ for all weeks prior to week 8 and similarly expand $D_\text{test}$ to include information on interns from $D_\text{cal}$ after week 8. The results have been averaged over all interns and decision points in $D_\text{test}$ (4 weeks worth of data; week 9-12) and are provided in Table \ref{tab:ihscomp_predout}. Our observations are similar to the predicting downwards case, where the linear model seems to have better performance than the other machine learning models. Moreover, these intervals are wider than those obtained while predicting downward, highlighting the difficulties associated with predictions at future unobserved time points.

As a reminder, performance is evaluated using average length and pseudo-outcome coverage, which we are considering a surrogate for actual ITE coverage. 

\begin{table}[ht]
\caption{Performance of ML models when creating prediction intervals for IHS (2020) data when predicting outward. Results averaged over all individuals and decision points in the test data. Best performing model's average length bolded. \emph{Abbreviations}: PCov, Pseudo-outcome coverage; AL, Average length; LIN, Linear model with $L1$ penalty (LASSO); NN, Neural network; RF, Random forest.\\}
\centering
{\small
\begin{tabular}{@{}lllllll@{}}
\toprule
\multirow{2}{*}{} & \multicolumn{4}{c}{Outcome}                                                      \\ \cmidrule(l){2-5} 
                    Model   & \multicolumn{2}{l}{Step count} & \multicolumn{2}{l}{Sleep}  \\ \cmidrule(r){2-5}
                       & PCov         & AL          & PCov         & AL             \\
                       \cmidrule(r){1-5}
LIN                    & 94.31       & \textbf{5163.40}       & 94.66       & \textbf{141.57}             \\
NN                     & 92.55       & 5773.49       & 94.30       & 194.54         \\
RF                     & 93.14       & 5282.42       & 94.97       & 155.64       \\ \bottomrule 
\end{tabular}
}

\label{tab:ihscomp_predout}
% \end{centering}
\end{table}

\newpage

\subsection{Notation and abbreviation guide}
\label{app:notation}

For convenience, we summarize some key notation used throughout the paper.

\begin{itemize}
    \item $N$: total number of individuals 
    \item $T$: total number of decision points
    \item $[k] = \{  1,2,...,k \}$
    \item $(i,j)$: subscripts for the $i^\text{th}$ individual at the $j^\text{th}$ decision point
    \item $A_{ij}$: binary action (treatment)
    \item  $Y_{ij}$: observed outcome
    \item $X_{ij}$: subset of the observed characteristics until and including the $j^{th}$ decision point
    \item $\bar{A}_{ij}$: set of all actions \textbf{until and not including} decision point $j$
    \item $Y_{ij}(\bar{A}_{ij}, 0)$ and $Y_{ij}(\bar{A}_{ij}, 1)$: the potential outcomes if $A_{ij} = 0$ or $1$ respectively. 
    \item $\mathcal{D}$: data format $=(X_{ij}, A_{ij}, Y_{ij}) \in \mathcal{X} \times \{0,1\} \times \mathbb{R}, ~ \forall ~ (i,j) \in [N] \times [T]$,  where $A_{ij} \in \{0,1\}$
    \item $(X_{I,J}, Y_{I,J})$:  test unit
    \item $\hat{C}(X_{I,J})$: prediction interval/set
    \item $\pi(X_{ij}, \bar{A}_{ij})$: propensity score $P(A_{ij} = 1 | X_{ij}, \bar{A}_{ij})$
    \item $\mu_a$: conditional average outcome for fixed treatment~$a \in \{0,1\}$
    \item $\varphi = (\mu_0, \mu_1, \pi)$: Nuisance parameters
    \item $\widetilde{Y}_\varphi$: pseudo-outcomes
    \item $D_\text{tr}$: training data
    \item $D_\varphi, D_\text{model}, D_\text{cal}$: 3 equal splits of $D_\text{tr}$
    \item $D_\text{test}$: test data
    \item $(\hat{q}_\text{lo}, \hat{q}_\text{hi})$: quantile regression models 
    \item $Z_{ij}$: pair   $(X_{ij},\Tilde{Y}_{\varphi,ij})$
    \item $Z$: sequence $(Z_{1:N, 1:T}, Z_{I,J})$
    \item $Z^{(i,j)}$: swaps indices $(i,j)$ with $(I,J)$
    \item $w_{ij}$: weights
    \item $\overset{\sim}{w}_{ij}$: normalized weights
    \item $\mathcal{A} = (\hat{q}_\text{lo}, \hat{q}_\text{hi})$ 
    \item $\mathcal{K}= \left\{ [N] \times [T], (I,J)  \right\}$
    \item $\text{d}_\text{TV}$ or TVD: total variation distance
    \item $V_\varphi$: observed conformity scores
    \item $V^*$: oracle conformity scores
    \item $Y_\text{ij,ITE}$: true ITE = $Y_{ij}(\bar{A}_{ij},1) - Y_{ij}(\bar{A}_{ij},0)$
    \item $\alpha$: pre-specified significance level
    \item  $V_\varphi \geq_{(1)} V^*$: First order stochastic dominance (FOSD)
    \item  $V_\varphi \leq_{(2)} V^*$: Second order stochastic dominance (SOSD)
    \item $V_\varphi \geq_{mcx} V^*$: Monotone convex dominance (MCX)
    \item $\psi$: decay parameter in weighting strategies
\end{itemize}

We provide a list of all abbreviations used in the paper.

\begin{itemize}
    \item ITE: Individual Treatment Effect
    \item mHealth: Mobile Health
    \item MRT: Micro-randomized Trials
    \item E: Equal weights method
    \item D: Decaying weights method
    \item DSQ: Decaying squared weights method
    \item DRT: Decaying root weights method
    \item LASSO: Linear Absolute Shrinkage and Selection Operator
    \item ML: Machine Learning
    \item Cov: Coverage
    \item PCov: Pseudo-outcome coverage
    \item AL: Average length
    \item LIN: Linear model
    \item NN: Neural Network
    \item RF: Random Forest
    \item IHS: Intern Health Study
\end{itemize}

\end{document}